\documentclass[%
reprint,
 superscriptaddress,
nofootinbib,
 amsmath,amssymb,
pra,
floatfix,
]{revtex4-2}

\usepackage{mathtools}

\usepackage{subfig}
\usepackage{graphicx}
\usepackage{dcolumn}
\usepackage{bm}
\usepackage[hidelinks,colorlinks=true,linkcolor=blue,citecolor=blue]{hyperref}
\usepackage{braket}

\begin{document}

\preprint{APS/123-QED}

\title{Large Spin Stern-Gerlach Interferometry for Gravitational Entanglement}

\author{Lorenzo Braccini} \email{lorenzo.braccini.18@ucl.ac.uk}
\affiliation{Department of Physics and Astronomy, University College London, Gower Street, WC1E 6BT London, United Kingdom}
\author{Martine Schut}
\affiliation{Van Swinderen Institute for Particle Physics and Gravity, University of Groningen, 9747 AG Groningen, Netherlands}

\author{Alessio Serafini}
\affiliation{Department of Physics and Astronomy, University College London, Gower Street, WC1E 6BT London, United Kingdom}

\author{Anupam Mazumdar}
\affiliation{Van Swinderen Institute for Particle Physics and Gravity, University of Groningen, 9747 AG Groningen, Netherlands}
\author{Sougato Bose}
\affiliation{Department of Physics and Astronomy, University College London, Gower Street, WC1E 6BT London, United Kingdom}

\date{\today}

\begin{abstract}
Recently, there has been a proposal to test the quantum nature of gravity in the laboratory by witnessing the growth of entanglement between two masses in spatial quantum superpositions. The required superpositions can be created via Stern-Gerlach interferometers, which couple an embedded spin qubit quantum state to the spatial dynamics of each mass. The masses would entangle only if gravity is quantum in nature. Here, we generalise the experiment to an arbitrary spin $j$ or equivalently to an ensemble of uniformly coupled spins. We first exemplify how to create a generalized Stern-Gerlach interferometer, which splits the mass into $2j+1$ trajectories.
This shows that a controlled protocol can be formulated to encode the amplitudes of any spin state to a spatial superposition. Secondly, two masses in spatial superpositions of the above form are left to interact via gravity, and the entanglement is computed. Different families of initial spin states are varied to find the optimal spin state that maximizes the entanglement. We conclude that larger spins can offer a modest advantage in enhancing gravity-induced entanglement.
\end{abstract}

\maketitle


\section{\label{sec:introduction}Introduction}

In the last decades, the search for a unifying theory of general relativity and quantum mechanics has been central in discussions of theoretical physics. 
Historically, a crucial difficulty faced by the Quantum Gravity (QG) community has been the absence of related phenomena detectable in a laboratory setting. 
In the low energy regime, where most of the experiments are conceived, it is tough to test directly the carrier of the gravitational force, known as the massless spin-2 graviton~\cite{dyson_is_2013, rothman_can_2006}. 
In the high energy regime (\mbox{$M_{p}\sim 10^{19}$}~GeV), it is inconceivable to probe the time scale ($t_p\sim 10^{-44}$ s) and length scale ($\ell_p\sim 10^{-35}$ m) of spacetime where it is believed that actual non-perturbative QG effects would be relevant.

Recently, a promising proposal has emerged which will enable us to witness the quantum coherent nature of gravitational superpositions while still staying within the weak field regime. 
The proposed experiment uses a quantum spatial superposition of mesoscopic objects (say, via a Stern-Gerlach interferometer \cite{keil_stern_2021}).  The state describing the whole system is assumed to be initially in a product state, and only the gravitational interaction is present. The proposed experiment is based on the fact that local operation and classical communication cannot yield entanglement~\cite{bennett_mixed_1996, horodecki_quantum_2009}. Thus, if gravity - the sole channel between the masses - is inherently quantum, it would entangle two test masses in a quantum superposition, while a classical natured gravity cannot. Measurement of entanglement would genuinely witness the quantum nature of gravity. This scheme can be dubbed as Quantum Gravity-induced Entanglement of Masses (QGEM)~\cite{bose_matter_2016,bose_spin_2017,marletto_gravitationally-induced_2017}.

The lowest order interaction between the masses is Newtonian, which in a perturbative quantum theory of gravity results from the interaction between the masses through virtual graviton exchanges. 
That this process can provide the quantum communication necessary to entangle the two objects as highlighted in Ref.~\cite{marshman_locality_2020,bose_mechanism_2022}. In this regard witnessing the entanglement between the two test masses due to their Newtonian interaction is similar in spirit to Bell's test of quantum mechanics~\cite{hensen_loophole_2015}. Another way of justifying the protocol is the observation that entanglement generation requires an operator valued interaction between the masses which is not possible with a classical mediator, as was shown explicitly by quantizing the spin-2 graviton around a Minkowski background~\cite{bose_mechanism_2022}.
The idea that entanglement detection would witness QG is supported by several independent works, namely that it is an evidence of the quantum superposition of geometries~\cite{christodoulou_possibility_2019}, by the relation between different degrees of freedom of the gravitational field~\cite{belenchia_quantum_2018, danielson_gravitationally_2022, carney_newton_2022}, path integral calculations~\cite{christodoulou_locally_2022}, generalized probabilistic theories~\cite{galley_no-go_2022}, quantum reference frames~\cite{christodoulou_gravity_2022}, and canonical quantization of gravity~\cite{chen_quantum_2022,vinckers_quantum_2023,Chakraborty:2023kel}. 

This class of proposed experiments have many other potentialities to study fundamental questions: spin-2 character of the exchanged graviton via light-matter entanglement (a quantum counterpart of the bending of light)~\cite{biswas_gravitational_2022}, discreteness of time~\cite{christodoulou_possibility_2020, christodoulou_experiment_2022}, the weak equivalence principle~\cite{bose_entanglement_2023}, fifth force~\cite{barker_entanglement_2022}, massive graviton~\cite{elahi_probing_2023}, and non-local gravitational interactions arising from string theory~\cite{vinckers_quantum_2023}.
Alternative apparatus, such as optomechanical systems~\cite{al_optomechanical_2018,miao_quantum_2020,matsumura_gravity-induced_2020} and quantum gasses~\cite{haine_searching_2021}, have been studied in the literature, showing both potentialities and difficulties for detecting entanglement. Non-Gaussianity of Bose-Einstein condensates~\cite{howl_non-gaussianity_2021} may also be proof of QG. Instead of witnessing gravitational entanglement, a recent proposal has been developed that emphasizes testing the quantum nature of gravity via the measurement-induced disturbance on a quantum system~\cite{hanif_testing_2023}.

The aim of our paper is to investigate whether large spins embedded in nanocrystals could increase the entanglement generated by the gravitational interaction. This is motivated by several considerations. First of all, in the quantum sensing community, it has been noted that large spins (and spins ensembles) are excellent sensors~\cite{degen_quantum_2017}. Squeezed state and superposition of coherent states of the spin system reach the fundamental Heisenberg limit~\cite{ ma_quantum_2011, maleki_quantum_2021}. 
Thus, large spins are a promising system for sensing the weak entanglement generated by the gravitational interaction between two mesoscopic masses. Secondly, large spins in ferromagnetic nanoparticles could be used to create large superpositions~\cite{rahman_large_2019}. Thirdly, so far, the spin qubit based gravitational entanglement experiments have been formulated in terms of a single Nitrogen Vacancy (NV) centre in a mesoscopic diamond crystal. However, it is an experimental challenge to create a single NV-center in a nano-diamond, as multiple NVs (spin qubits) may be present. In cases when the simplifying assumption that the coupling with the magnetic gradient is the same for each spin qubit is valid, and when we can neglect the spin-spin coupling, the system can be treated as a large effective spin.

We consider a large spin undergoing a spatial splitting: we will call this a \textit{ ``Generalized Stern-Gerlach"} (GSG) interferometer. 
As the first part of the paper, we describe the dynamics of this process and find that the superposition distance spreads proportionally to total angular momentum $j$, while the splitting time remains constant. The mass is superposed in $2j + 1$ coherent states, i.e. semi-classical paths. 
The protocol exemplifies a way to encode any spin state to a spatial superposition, fully capturing the ability of large spins in spatial-dependent protocols. As the second part, two such states, each with 
$2j + 1$ semi-classical paths generated due to spin-motion interaction (for example, as generated in adjacent GSGs, but could be otherwise), is let interact via gravity. The gravitational entanglement generated is computed and compared for various values of spins. 

The paper is organized as follows. In Sec.~\ref{sec:generalized}, the GSG interferometer is presented, deriving the splitting and recombining dynamics that create the one-loop interferometer. We find that GSG is a map from any spin state to a spatial superposition. 
Sec.~\ref{sec:spins_states} introduces the spins states that are used in this study (coherent spin states, a superposition of coherence spin states and squeezed spin states). 
In Sec.~\ref{sec:interferometers_geometries}, two different two-interferometer geometries are presented. 
The dynamics of these set-ups under gravitational interaction is described in Sec.~\ref{sec:gravitation}, and the optimal spin states that maximize the gravitational entanglement are numerically computed in Sec.~\ref{sec:entropy}. 
A suitable witness and the influence of decoherence is discussed in Sec.~\ref{sec:witness} and Sec.~\ref{sec:decoherence}, respectively. 
The concluding remarks are given in Sec.~\ref{sec:conclusion}.

\section{\label{sec:generalized}Generalized Stern-Gerlach Interferometer with Large Spins}

Let us consider a large spin $j$ embedded in a mass~$M$. The spin Hilbert space $ \mathbb{C}^{2 j + 1}$ is spanned in the Dicke basis $\{ \ket{j, m} \}$ with $m \in \left[-j, j\right]$, which are the  eigenvectors of the $J_z$ operator, where $z$ is the quantization axis. 
The $j$-dependency of the quantum states will not be carried forward, as $j$ is assumed to be constant over the experiment.
Such a system can physically arise either from a single large spin $j$ or an ensemble of $n$ spin–1/2 particles, with $n = j/2$, embedded in a nanocrystal. The latter requires that the spin ensemble is permutationally invariant, i.e. each spin couples uniformly with the external forces, and no spin-spin interaction is present~\cite{dicke_coherence_1954, schliemann_coherent_2015}. While this may not be always an accurate assumption, it is sometimes used for spins ensembles in a nano-crystal \cite{wang_giant_2022}, and we are going to make this simplifying assumption.

We consider a protocol that can be described in $6$ stages as follows:

\vspace{0.2cm}
(1) \textit{System Initiation:} The system is initially prepared in the ground state for both spin and position degrees of freedom, i.e. $\ket{\psi} = \ket{\psi_S} \otimes \ket{\psi_x} = \ket{m=- j} \otimes \ket{0}$.
For the position degree of freedom, we assume that the ground state can be achieved by usual feedback cooling of the mass in a trap~\cite{delic_cooling_2020}. 
The spin state can be created by applying a strong magnetic field in the direction of the quantization axis. Subsequently, both the trapping mechanism and the magnetic field are turned off. 

\vspace{0.2cm}
(2)  \textit{Spin State Preparation}: A desired spin state can be created by applying secondary magnetic fields or by using microwave and radio-frequency field pulses. Generally, the spin state will take the form:
\begin{equation}
    \ket{\psi_S} = \sum_{m = -j}^j c_m \ket{m} \, ,
\end{equation}
where $\{ c_m \}$ are the coefficients in the Dicke basis. A review of the spin states considered in this work is given in Sec~\ref{sec:spins_states}, while the following dynamics is kept in full generality.

\vspace{0.2cm}
(3) \textit{Splitting Process:} A uniform magnetic gradient $ \left( \partial_x B \right)$ is applied to the mass. We will assume that there is no external rotation of the mass~\cite{japha_quantum_2023,ma_torque_2021}. The splitting Hamiltonian is then given by: 
\begin{equation}
\label{splitting_hamiltonian}
    H =  \hbar \omega_M a^\dag a - \hbar g J_z \left( a + a^\dag \right) \, .
\end{equation}
The first term is that of the quantum harmonic oscillator in which the mass is trapped, with frequency $\omega_M$ and ladder operator $a$. The Stern-Gerlach spin-position coupling is the second term, where $J_z$ is the $z$-spin operator and the coupling constant for the mass $M$ is given by
\begin{equation}
\label{eq:coupling}
 g =  \tilde{g} \mu_B \sqrt{\frac{1}{2 \hbar M  \omega_M}}  \left( \partial_x B \right) \, ,
\end{equation} 
where $\mu_B$ is the Bohr magneton and $\tilde{g}$ is the Landé g-factor. 

It has to be noted that this work is the first theoretical description of GSG interferometry, and does not aim to resolve well-known experimental challenges of creating a mesoscopic superposition with the use of Stern-Gerlach interferometery. As pointed out in Ref.~\cite{pedernales_motional_2020, marshman_constructing_2022,Zhou:2022frl}, the diamagnetic response of the mass – a nanodiamond with a single NV center – placed in the magnetic gradient significantly increases the splitting time, representing one of the greatest experimental issues\footnote{However, once the initial superposition is created of decent splitting, the diamagnetic induced potential can also aid to create a larger superposition with the help of wires,  see~\cite{Zhou:2022jug,Zhou:2022epb}.}. We find that this challenge is also present for large spins, or multiple NV centers making an effective large spin; in Appendix~\ref{app:diamagnetic}, the Hamiltonian in Eq.~(\ref{splitting_hamiltonian}) is derived by considering the diamagnetic response of the mesoscopic mass and a linear magnetic gradient. Here we proceed by assuming that $\omega_M$ can be engineered to be independent of $ \left( \partial_x B \right)$. If this assumption is not made, all our results will still be valid except occurring over much longer time scales as in the spin-1/2 case. In essence, the assumption of the independence of $\omega_M$ to $ \left( \partial_x B \right)$ is difficult to fulfill, without being able to nullify the effect of diamagnetism. Hence, this assumption is equivalent to be able to engineer a composite material with extremely low magnetic susceptibility. We are making this simplifying assumption herewith in order to make progress to be able to compare the efficacy of large vs small spins in equal footing in generating gravitationally induced entanglement. 

The unitary evolution and the time evolution of the state resulting from the Hamiltonian in Eq.~(\ref{splitting_hamiltonian}) are derived in Appendix~\ref{app:unitary_evolution}. It follows that the quantum state at time $t$ is given by:
\begin{equation}
\label{quantum_state_evolution}
     \Ket{\psi(t)} = \sum_{m=-j}^{j} c_m e^{ i  \frac{g^2}{\omega_M^2} m^2 \left(\omega_M  t - \sin(\omega_M t)\right) }   \ket{ m}  \otimes  \Ket{\alpha_m (t)}\, ,
\end{equation}
where the position coherent states\footnote{Mathematically, position coherent states are defined by the application of a displacement operator to the vacuum state $\ket{\alpha} = \mathcal{D} (\alpha) \ket{0} = e^{\alpha^\dag a} \ket{\alpha}$. They are eigenstates of the annihilation operator ($a \ket{\alpha} = \alpha \ket{\alpha}$) and Gaussian states in both position and momentum representations. They form an over-complete basis, i.e. $\int_\mathcal{C} \ket{\alpha} \bra{\alpha} d^2 \alpha $ and they are states of minimum uncertainty.}  have parameters:
\begin{equation}
    \alpha_m (t) = m  \frac{g}{\omega_M} \left(1 - e^{-i \omega_M t}\right) \, .
\end{equation}
Under the evolution, the spin and position degrees of freedom entangle -- represented by the $m$-dependency in $\Ket{\alpha_m (t)}$ state of Eq.~(\ref{quantum_state_evolution}) --, and the mass' position degree of freedom branches out over $2j+1$ paths, i.e. coherent state trajectories. Each path has minimum uncertainty; its position and momentum expectation values are given by: 
\begin{align}
        \braket{x_m (t)} &= m  \frac{g}{\omega_M} \sqrt{\frac{2 \hbar}{M \omega_M}}  \left(1 - \cos(\omega_M t) \right) \label{eq:x_exp_val}, \\
        \braket{p_m (t)} &= m g \sqrt{\frac{2 \hbar M }{\omega_M} } \sin(\omega_M t) \, .
\end{align} 

\begin{figure}
    \includegraphics[width=0.45\textwidth]{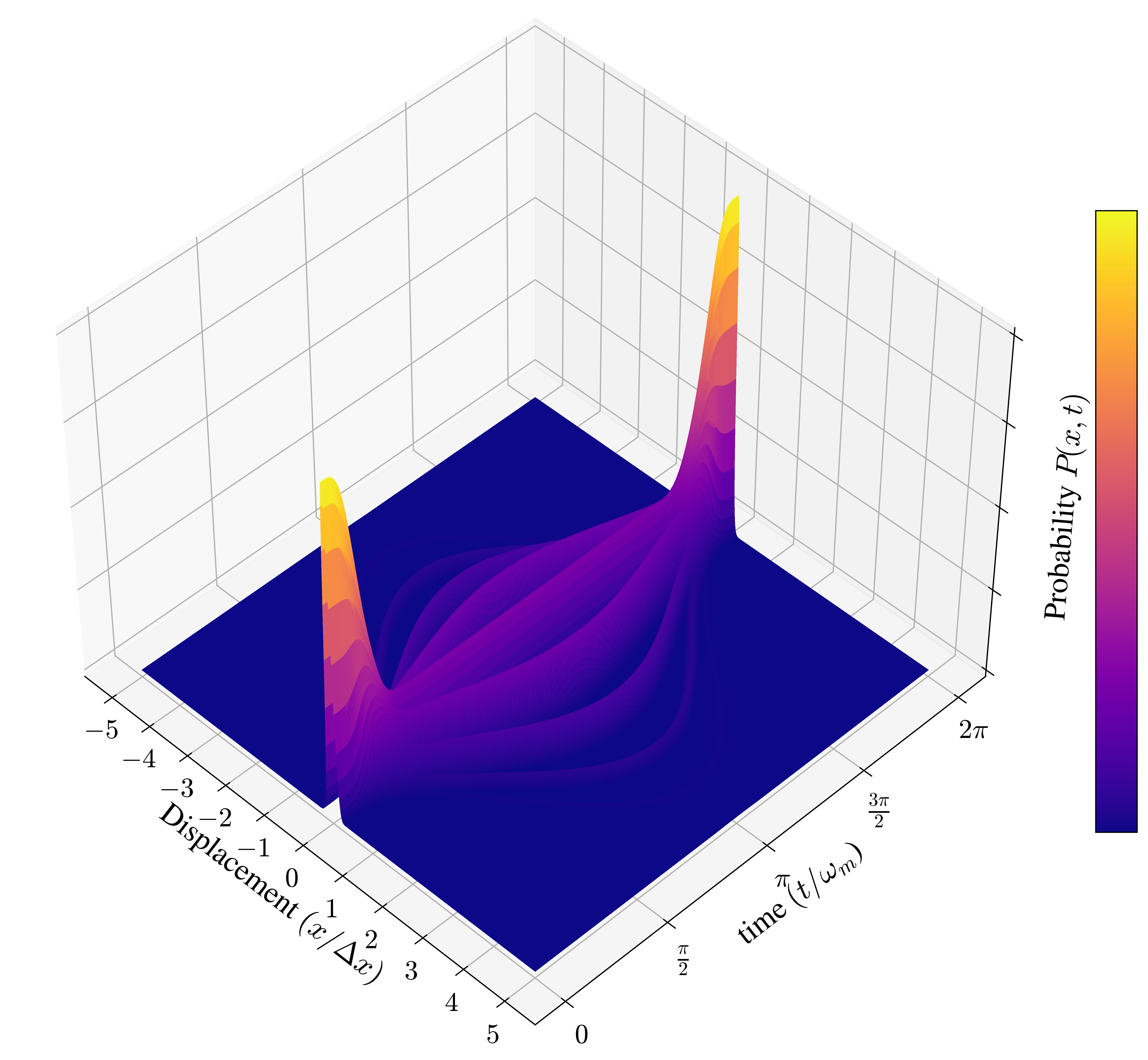}
    \caption{Probability distribution $P(x,t)$ as a function of time (in frequency unit $\omega_M$) in the position representation (in $\Delta x$ units) of a Coherent Spin State $\Ket{j=5; \phi = 0, \theta = \pi/2}$ in a GSG Interferomenter. The mass undergoes the splitting and recombining process in $11$ semi-classical paths, weighted by the probability distribution of the $m$-states.\label{plt:prob_t_x}}
\end{figure}

Let us define the spin probability distribution $P_s(m) := \left| \braket{\psi_S|m} \right| ^ 2 = |c_m|^2$. We denote a normal distribution with expectation value $\mu$ and standard deviation $\sigma$ as $\mathcal{N} \left(\mu, \sigma^2 \right)$. 
The probability of finding the mass at $x$ at time $t$ in the position representation~is:
\begin{equation}
    P\left(x,t\right) = \sum_{m = -j}^{j} P_s(m) \mathcal{N} \left(\braket{x_m (t)}, \sigma_{x_m}^2 = \frac{\hbar} {M \omega_M} \right) \, ,
\end{equation}
which represents a sum of Gaussians, weighted by the discrete spin probability distribution $P_s(m)$. An example of $ P\left(x,t\right) $ is given in Fig.~\ref{plt:prob_t_x}.

\vspace{0.2cm}
(4) \textit{Maximum Displacement and Interaction:}  
The maximum displacement between adjacent paths is achieved at the splitting time $t_s := \pi/\omega_M$, and takes the value:
\begin{equation}\label{eq:deltax_MS}
    \Delta x := \braket{x_{m+1}\left( t_s \right)} - \braket{x_m\left( t_s \right)}  = 2 \sqrt{\frac{2 \hbar}{M \omega_M}} \frac{g}{\omega_M} \, ,
\end{equation}
This splitting is independent on which adjacent paths $m$ and $m+1$ are taken, as the distance between each adjacent path is the same.

The spatial superposition spreads over the total range
\begin{equation}
    \Delta D := \braket{x_{j}\left( t_s \right)} -\braket{ x_{-j}\left( t_s \right)} = 2 j \Delta x \, ,
\end{equation}
i.e. it scales linearly with the total angular momentum, while the splitting time is independent of $j$, promising a larger total splitting within the same $t_s$ for higher $j$.

At the time $t_s$, the magnetic gradient is switched off and the system is described by the quantum state:
\begin{equation}
\label{eq:split_state}
     \Ket{\psi(t_s )} =  \sum_{m=-j}^{j} c_m e^{ i \pi \frac{g^2}{\omega_M^2} m^2}  \ket{m} \otimes \Ket{ \alpha_m = m \Delta x } \, .
\end{equation}
During this stage, the system can interact with external forces, such as gravity, and, as we will see, generate entanglement.

\vspace{0.2cm}
(5) \textit{Recombination Process:} Assuming that during the interaction the masses were not displaced
, the state can be recombined by applying a magnetic gradient with the same profile as during the \textit{splitting process} stage. After a time $t_s$ the mass is brought back to the center of the trap and the quantum state is give by:
\begin{equation}
\label{eq:final_state}
     \Ket{\psi(2 t_s)} = \left( \sum_{m=-j}^{j} c_m e^{ i 2 \pi \frac{g^2}{\omega_M^2} m^2}  \ket{m} \right) \otimes \Ket{0} \, ,
\end{equation}
where we assumed, for now, that no interaction was present during the previous stage. 
The recombining magnetic field has to be, with extreme precision, identical to the splitting one. 
Small differences between the two fields would lead to the Humpty-Dumpty effect ~\cite{englert_is_1988}, as the recombination process would be required to be a perfect time reversal of the splitting. 
Recently, a full-loop has been achieved in experiments~\cite{keil_stern_2021,amit_T_2019,margalit_realization_2020}, in the context of atom interferometry~\cite{machluf_coherent_2013}.

\vspace{0.2cm}
(6) \textit{Measurements:} One can note that there is no spatial dependency Eq.~(\ref{eq:final_state}): the spatial degrees of freedom are disentangled from the spin ones and the mass is  placed back to the center of the trap. the position component of the wavefunction can be trivially traced out, and, thus, general spin measurements can be carried out on the spin embedded in the mass.

\section{\label{sec:spins_states} Families of Spin States}

\begin{figure*}
\centering
  \subfloat[Spin Coherent State]{\includegraphics[width=0.33\textwidth]{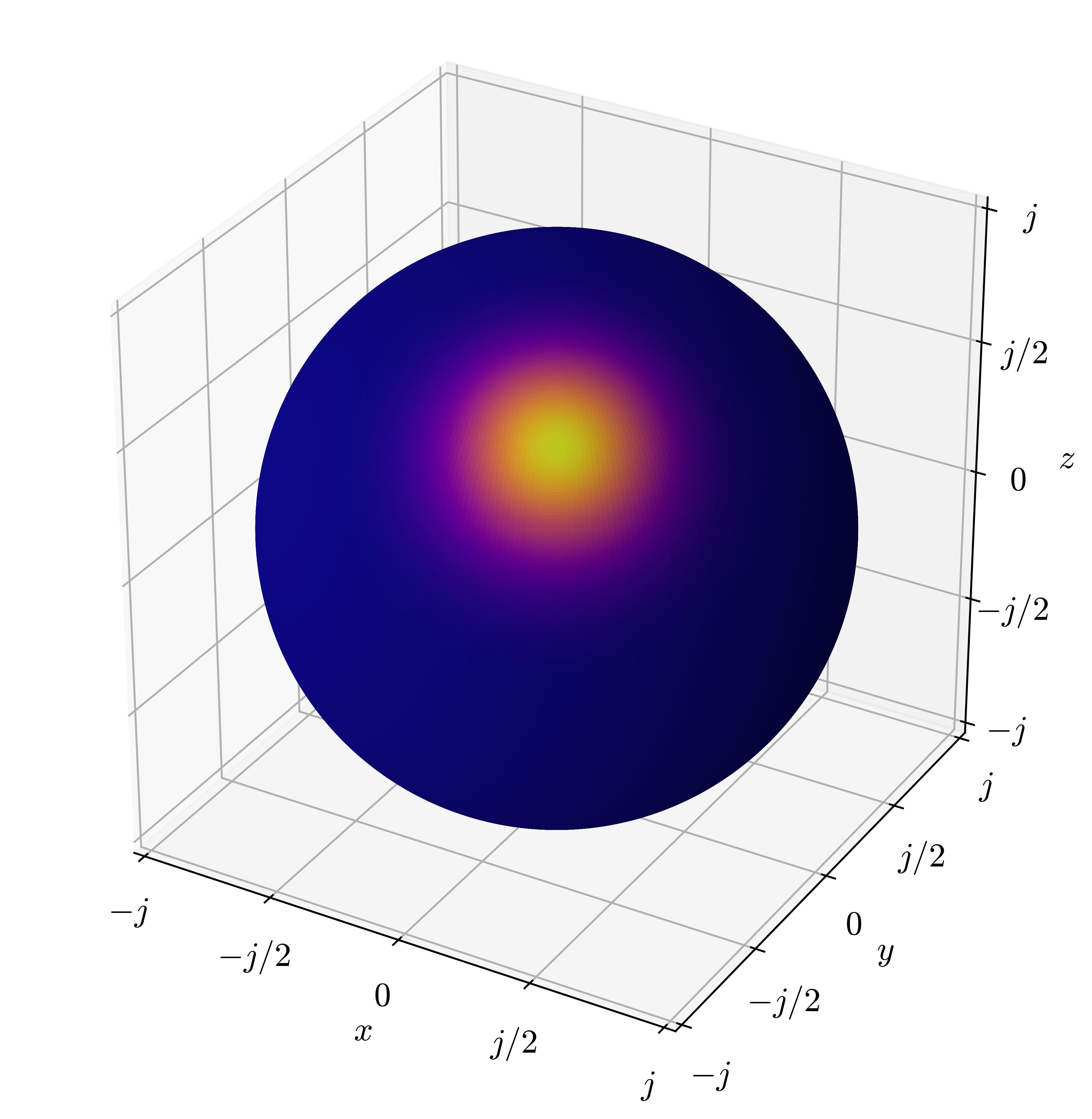}}
  \hfill
  \subfloat[Superposition of Two CSS]{\includegraphics[width=0.33\textwidth]{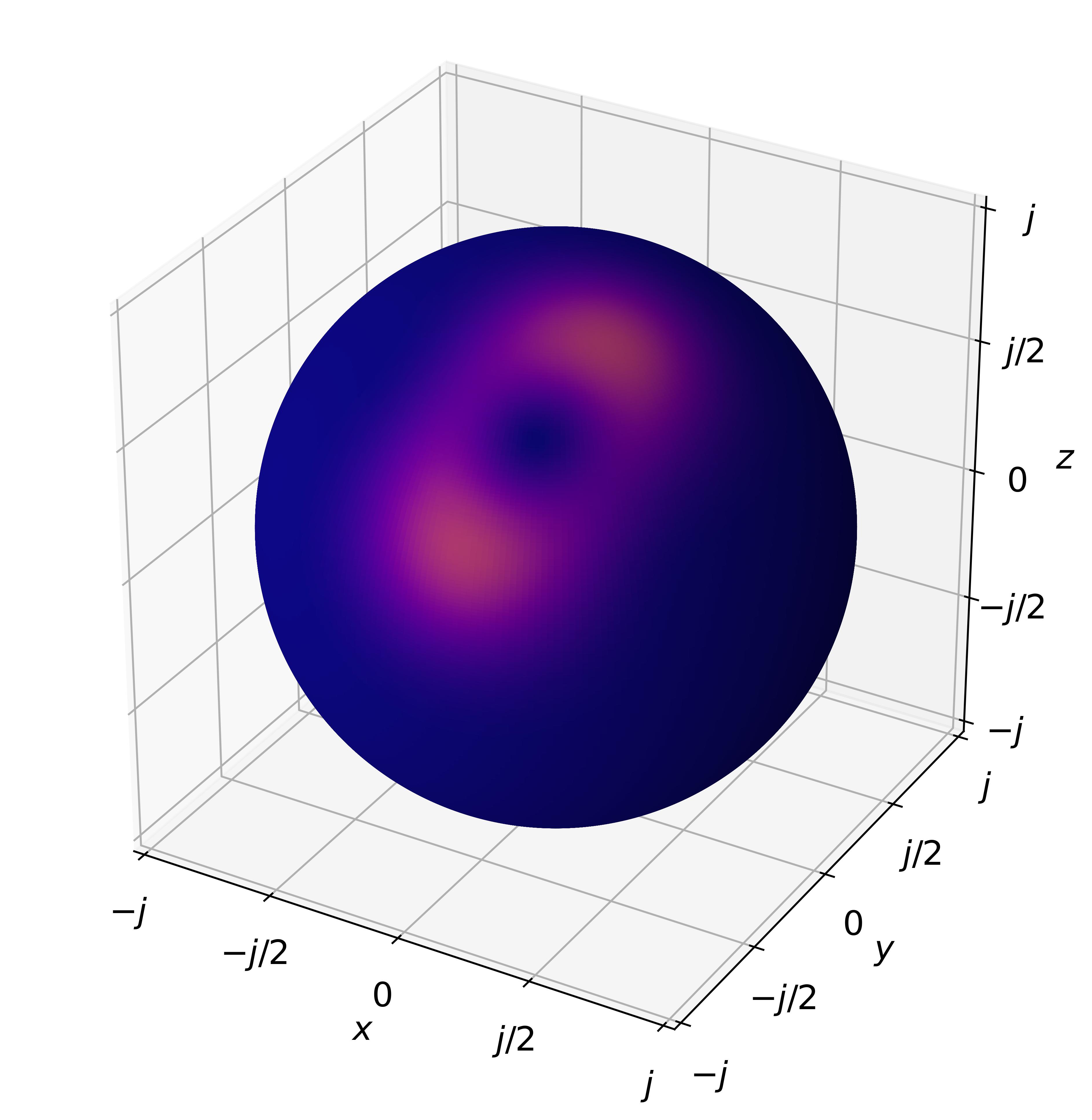}}
    \hfill
  \subfloat[Squeezed Spin State]{\includegraphics[width=0.33\textwidth]{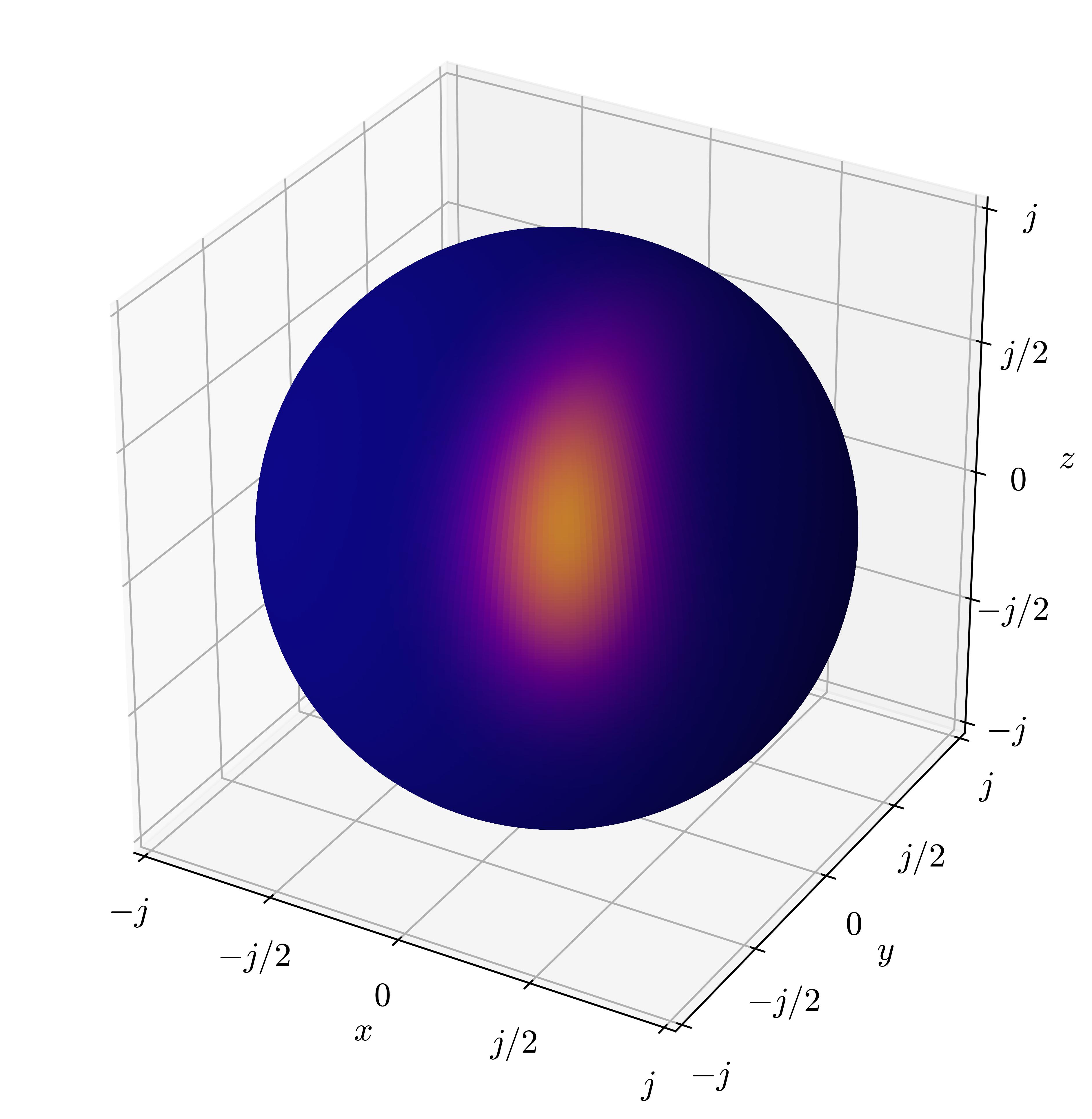}}
  \caption[width=\textwidth]{Examples of Husimi-Q quasiprobability distributions of (a) a Spin Coherent State, (b) a superposition of two CSS, and (c) a Spin Squeezed State. The axis are the spin axis with spin m from -j to j. }
    \label{plt:spin_states}
\end{figure*} 

The GSG interferometer represents a map from a spin state to an entangled superposition of spin and position coherent states.
The spins states $\ket{\psi_S}$ used in this study are: Coherent Spin States (CSS), a superposition of CSS, and Squeezed Spin State (SSS). In this Section a brief overview of the aforementioned states is presented. 
\vspace{-0.5cm}
\subsection{Coherent Spin States}
\vspace{-0.2cm}
A position coherent state of a quantum harmonic oscillator is a semi-classical state describing a translation of the vacuum state. 
Similarly, a CSS is defined as the state arising from an arbitrary rotation of the spin ground state, $\ket{-j}$\footnote{The CSS can be written in terms of Euler angles $\ket{\alpha, \beta, \gamma} := \mathcal{R}\left(\alpha, \beta, \gamma\right) \ket{m= j}$, where we used the Euler's angles $\alpha, \beta, \gamma$ and the arbitrary rotation $\mathcal{R}\left(\alpha, \beta, \gamma\right) \coloneqq e^{-i \alpha J_z}  e^{-i \beta J_y}  e^{-i \gamma J_z}$. }. 
This can be experimentally prepared by applying a magnetic field on an arbitrary axis or alternatively by applying a cavity field in resonance with the spin frequency. CSS states can be expanded on the Dicke basis in terms of  the Wigner D-Matrix. 
By defining the quantity $\mu = e^{i \phi} \tan{\theta/2}$, where $\theta$ is the azimuth angle with respect to the $z$-axis and $\phi$ is the $xy$-plane phase, a general CSS is:
\begin{align}\label{SCS_state}
\begin{split}
    \ket{\phi, \theta} &: =  \mathcal{N}e^{\mu J_-} \ket{m=- j} \\ 
    &= \mathcal{N} \sum_{m =-j}^{j} \mu^{j+m} \sqrt{\frac{2j!}{(j+m)! (j-m)!}} \ket{m} \, ,
\end{split}
\end{align}
where $J_-$ is the spin lowering operator and the normalization is given by~$P_s \approx \mathcal{N} = \left( 1 - \left|\mu\right|^2 \right)^{-j}$~\cite{radcliffe_properties_1971}. 
This parametrization has been chosen for its geometrical interpretation: the created CSS is a semi-classical spin state pointing in the  $\phi, \theta$ direction, i.e. the state with the maximum eigenvalue of the $\phi\theta-$spin projection operator \mbox{$J_{\phi, \theta} \ket{\phi, \theta} = j \ket{\phi, \theta}$}. 
The probability distribution in the $m$ representation is binomial and, thus, for large $j$, it approaches a Gaussian distribution $\mathcal{N} \left(\braket{\theta} = j \cos \theta,  \sigma_s^2 = j/2 \sin^2 \theta\right)$\cite{kofler_classical_2007}. 
Hence, in a GSG interferometer, the embedded CSS allows to manipulate the mass probability distribution. 
By varying the rotation angle $\theta$, the mass is centered at $ \Delta x \braket{\theta}$ with spread $\Delta x \sigma_s$\footnote{Given three orthogonal axis ($\alpha$, $\beta$, and $\gamma$), the uncertainty relation for a spin state is given by:
\begin{equation*}
    \left(\Delta J_\alpha \right)^2 \left(\Delta J_\beta \right)^2 \geq \frac{\left|\braket{J_\gamma} \right|^2}{4} \, .
\end{equation*}
Analogous to position coherent states, CSS are of minimum uncertainty $ (\Delta J_{\tilde{\alpha}} )^2 =  (\Delta  J_{\tilde{\beta}} )^2 = \hbar j/2$, where $\tilde{\alpha}$ and $\tilde{\beta}$ are orthogonal to the CSS axis (pointing in the $\theta$-$\phi$ direction)~\cite{kitagawa_squeezed_1993}.}. In the example given in Fig.~\ref{plt:prob_t_x}, the mass is centered and of maximum spread, as $\theta = \pi/2$.  
\vspace{-0.5cm}
\subsection{Superpositions of Coherent Spin States}
\vspace{-0.2cm}

A superposition of CSS can be considered, and we will limit ourselves to the simple case of a superposition of two states:
\begin{equation*}
\label{sup_CSS}
    \ket{\phi_1, \theta_1; \phi_2, \theta_2} \coloneqq \mathcal{\tilde{N}} \left( \ket{\phi_1, \theta_1} + \ket{ \phi_2, \theta_2}  \right),
\end{equation*}
where $\mathcal{\tilde{N}} $ is a normalization constant. Recently, such states has been of interest in the quantum-sensing community~\cite{maleki_quantum_2021} and the NOON\footnote{\mbox{$\ket{NOON} \equiv \frac{1}{\sqrt{2}} ( \ket{m = -j} + \ket{m= j}) = \frac{1}{\sqrt{2}}  ( \ket{\theta = 0} + \ket{\theta = \pi})$} with $\phi=0$.} state employed to create large mass superposition~\cite{rahman_large_2019}. 

\vspace{-0.5cm}
\subsection{Squeezed Spin States}
\vspace{-0.2cm}

Finally, the last spin states that will be considered are SSS: states with $ \left(\Delta J_\alpha \right)^2 < \frac{\left|\braket{J_\gamma} \right|}{2}$, with ($\alpha \neq \gamma \in (x,y,z)$. 
Contrary to the quantum harmonic oscillator case, for spin states there is not a  unique notion of the squeezing parameter, in literature several have been proposed~\cite{ma_quantum_2011}. 
Being agnostic on such an issue, we define these states from a state-preparation perspective. 
A SSS is created by the evolution of a CSS under the unitary evolution arising from non-linear Hamiltonians in the spin operators: the one-axis twisting Hamiltonian $H_{\text{OA}} \propto J_y^2$, which is experimentally easier to implement, and the two-axis twisting Hamiltonian $H_{\text{TA}} \propto J_+^2 - J_-^2$\cite{kitagawa_squeezed_1993}. 
For instance, we can define a one-axis twisted SSS as:
\begin{equation}
\label{SSS_state}
    \ket{\chi_{OAT}; \phi, \theta} \coloneqq e^{-i \chi_{OAT} J_y^2} \ket{\phi, \theta} \, ,
\end{equation}
and similarly for two-axis twisted SSS. 

The probability distribution of these states in the $m$ basis is sub-binomial~\cite{ma_quantum_2011}. Thus, with the use of a GSG, these states represents a controlled way to increase and decrease the spread of the weights of the paths.

In order to picture the aforementioned spin states $\ket{\psi_S}$, we compute the quasiprobability distribution known as the Husimi-Q function, defined according to $Q(\theta, \phi) \coloneqq \left| \braket{\theta, \phi| \psi_S} \right|^2 $. The Husimi-Q functions of each of the aforementioned spin states is presented in Fig.~\ref{plt:spin_states}. To conclude the section, CSS represents a 2-parameter family of states (with parameters $\theta$ and $\phi$), superposition of two CSSs are 4-parameter family ($\theta_i$ and $\phi_i$, with $i \in \{1,2\}$), and the SSS is a 3-parameter family ($\chi$, $\theta$ and $\phi$). These parameters can be varied to find the optimal states to maximize the detectable entanglement, for instance, generated by gravity. 

\vspace{-0.5cm}
\section{\label{sec:interferometers_geometries}Geometries for the Intereferometers}
\begin{figure}[!ht]
    \centering
    \includegraphics[width=8cm]{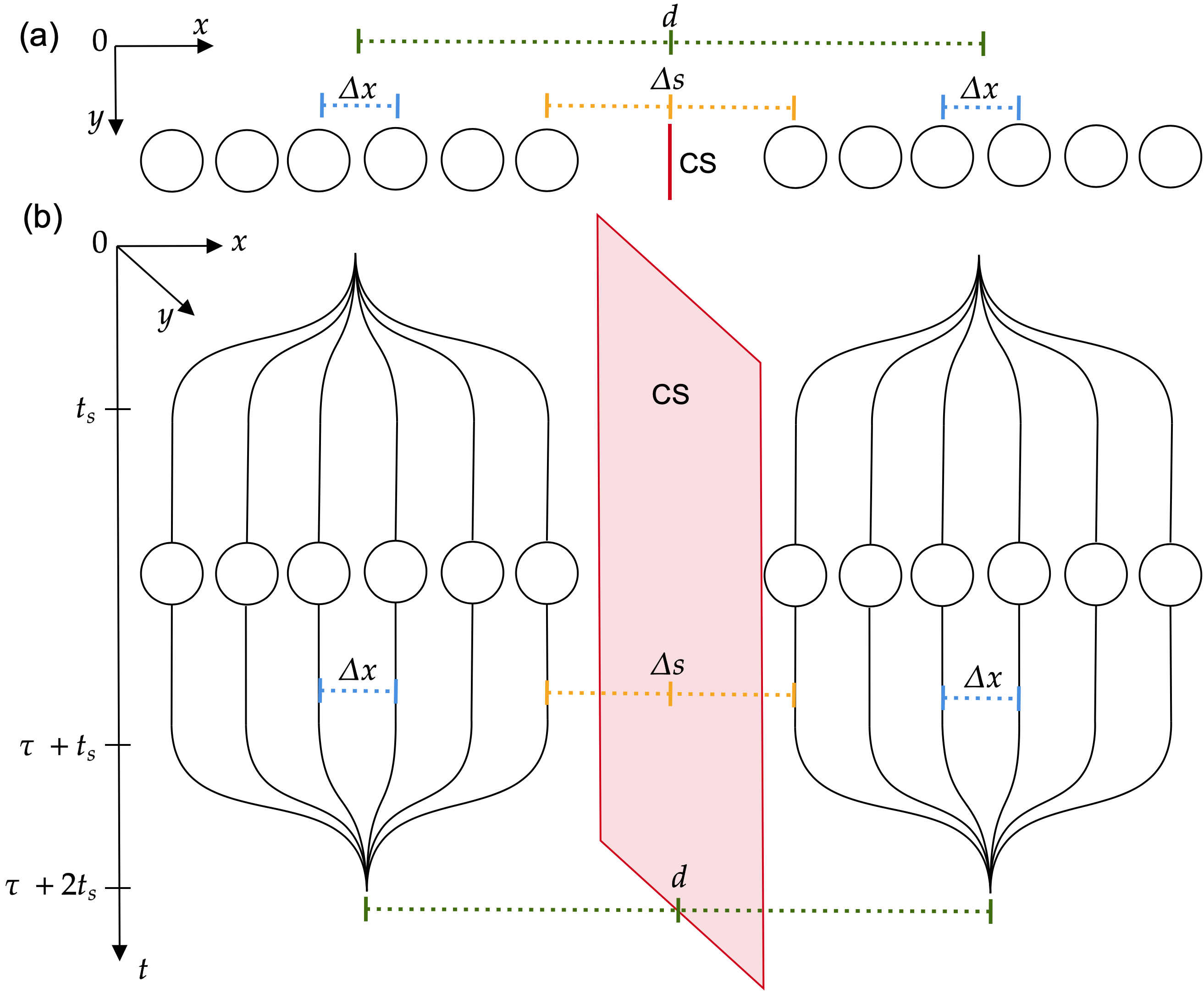}
    \caption{\label{linear_diagram}Linear set-up: two spins ($J=5/2$) at distance $d$ are spatially superposed and recombined on the $x$-axis via two GSG interferometers, with superposition distance $\Delta x$. The minimum distance between the two closest paths is $\Delta s = d - j \Delta x$. A \textit{Casimir shield} (CS) is placed between the masses and aligned with the $y$-axis.}
\end{figure}

\begin{figure}[!ht]
    \centering
    \includegraphics[width=8cm]{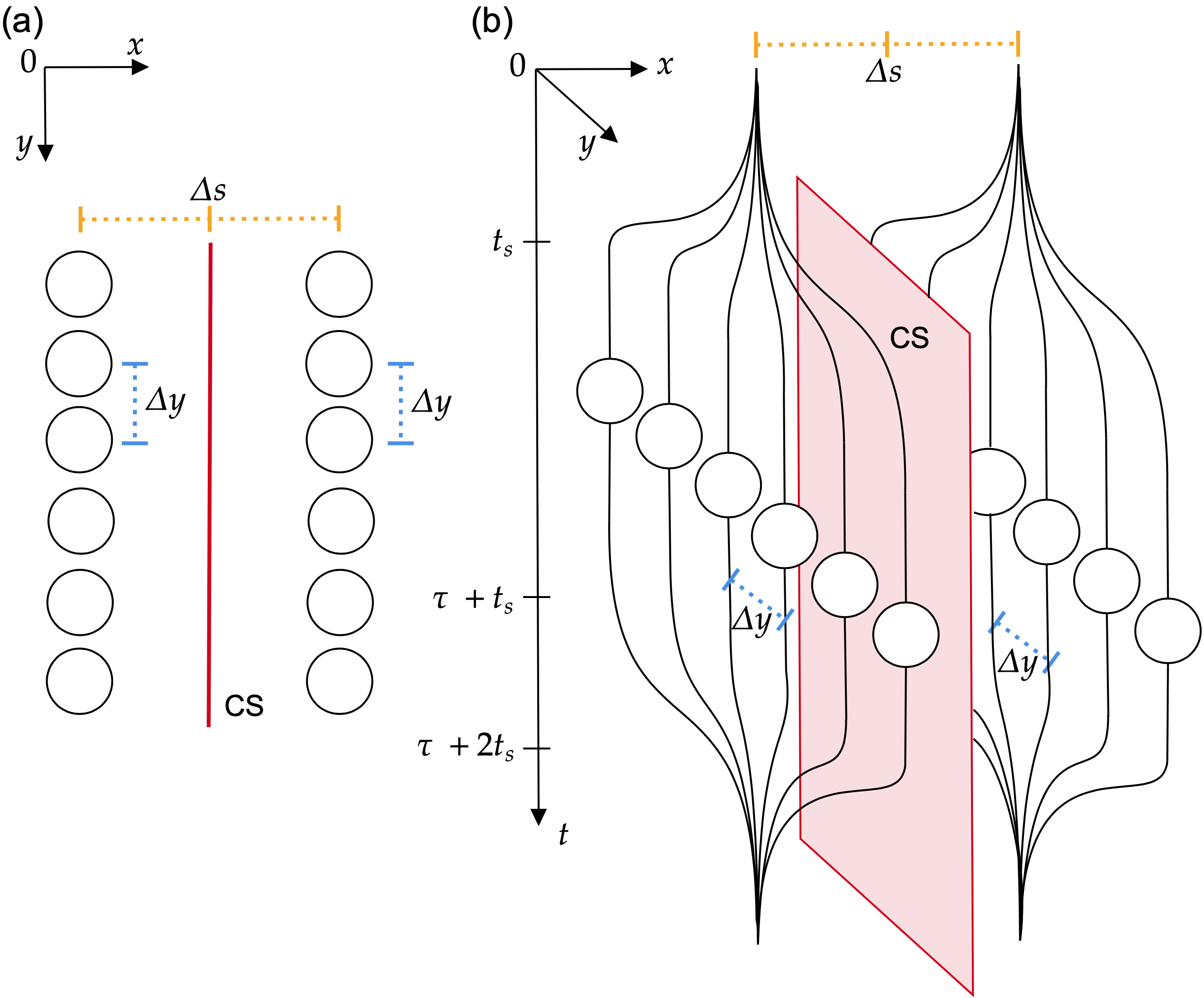}
    \caption{\label{parallel_diagram}Parallel set-up: two spins ($J=5/2$) at distance $\Delta s$ are spatially superposed and recombined on the $y$-axis via two GSG interferometers, with superposition distance $\Delta y$. The minimum distance between the two closest paths is $\Delta s = d/2$. A \textit{Casimir shield} (CS) is placed between the masses and aligned with the $y$-axis.}
\end{figure}

\begin{samepage}
In the original experimental protocol~\cite{bose_spin_2017}, two masses of \mbox{$M \sim \mathcal{O} (10^{-14}) {\rm Kg}$}, initially placed at distance \mbox{$\Delta s \sim {\cal O}(250){\rm \mu m}$}, are individually put in a spatial superposition of \mbox{$\Delta x\sim {\cal O}(200){\rm \mu m}$}. In order to witness QG, two GSG interferometers have to be placed at distance $\Delta s$, and let interacting via gravity. In this section, two geometries of the set-up will be presented (represented in Fig.~\ref{linear_diagram} and \ref{parallel_diagram}). 

The photon-induced Casimir-Polder (CP) interaction between two spheres~\cite{casimir_influence_1948,casimir_attraction_1948} is the main source of unwanted entanglement between the neutral masses. The CP potential is described by
\begin{equation*}
    V_{\text{CP}} (r) \approx - \frac{23 \hbar c}{4 \pi } \frac{R^6}{r^7} \left( \frac{\epsilon- 1}{\epsilon + 2}\right)^2 \; ,
\end{equation*}\end{samepage}
where $R$ is the radius of the mass, $\epsilon$ its dielectric constant, $r$ the distance between the two masses\footnote{The Casimir-Polder potential is given in Ref.~\cite{casimir_influence_1948}. 
We assume that the separation between the plate and the mass ($r$) is much larger than the radius of the mass ($R$), i.e. $r\gg R$.
We have furthermore assumed that the polarizability can be expressed as $R^3(\epsilon-1)/(\epsilon+2)$, which is known to be the case for e.g. diamond in a high vacuum~\cite{kim_static_2005}. The equation in this paper is found based on these assumptions, see also Ref.~\cite{van_de_kamp_quantum_2020}.}. 

Since the CP potential scales with $r^{-7}$, while the Newtonian potential scales with $r^{-1}$, this electromagnetic phenomenon restricted the aforementioned experimental parameters to a \textit{large} distance between the two interferometers ($\Delta s \sim {\cal O}(200){\rm \mu m}$), such that the gravitational-induced entanglement dominates at least ten times over the electromagnetic one. However, it was later realized that separating the two interferometers by a conducting plate ameliorates the parameter space tremendously~\cite{van_de_kamp_quantum_2020,schut_relaxation_2023}, reducing the separation distance within $\Delta s \sim {\cal O}(50){\rm \mu m}$. We will consider screening in our proposal, which will allow for a higher gravitational entanglement. 

Furthermore, following the insights ~\cite{nguyen_entanglement_2020,tilly_qudits_2021,schut_improving_2022}, two interferometric geometries will be used, namely, `linear' and `parallel' set-ups.
They are illustrated in Fig. \ref{linear_diagram} and Fig.~\ref{parallel_diagram}, respectively. 

\section{\label{sec:gravitation}Gravitational Entanglement}

As exemplified in Sec. \ref{sec:generalized}, it is possible to create an arbitrary spin-position quantum state of the form in Eq.~(\ref{eq:split_state}). We assume that, after the splitting, the two masses are in tensor product of these quantum states\footnote{Hence, we are neglecting entanglement generated during the splitting process. This was done to not commit to a specific trajectory (in our case a sinusoidal function), which may change with different splitting processes in experimental setting, and ensure to give a comparative analysis between different spins.}. The quantum state of the composite system is \mbox{$\ket{\Psi_{AB}(t_s)} = \Ket{\psi_A(t_s)} \otimes \Ket{\psi_B(t_s)}$}, where:
\begin{equation*}
     \Ket{\psi_I(t_s)} = \sum_{m=-j}^{j} c_m^{(I)} e^{ i \pi \frac{g^2}{\omega_M^2} m^2}  \ket{m} \otimes \Ket{ x_m^{(I)}(t_s) } \, ,    
\end{equation*}
whit $ x_m^{(I)}(t_s) = m \;\Delta x + x_0^{{(I)}}$, $I \in \{ A, B\}$. The position $x_0^A$ and $x_0^B$ are the centers of the traps of each GSG interferometer: in the linear set-up $x_0^A- x_0^B = d$, while in the case of parallel $x_0^A- x_0^B = \Delta s$.

The system is let to interact via gravity for a time~$\tau$. In the low energy regime of the experiment, the gravitational interaction can be approximated by the operator:
\begin{equation*}
\hat{H}_G = \frac{GM_AM_B}{\left| \hat{x}_A - \hat{x}_B \right|} \, ,
\end{equation*}
where $\hat{x}_{I}$ ($I=A,B$) are the position operators of each mass. 
We are going to use the approximation that the position coherent state representing each of the semi-classical paths is an eigenstate of the corresponding position operator, i.e.~$\hat{x}_I \ket{x_m^{(I)}} \approx  x_m^{(I)} \ket{x_m^{(I)}} $. 
This is motivated by the fact that the distances involved in the experiment are much greater than the spread of the position coherent state, $\braket{ x_m^{(I)}} \gg \sqrt{ \braket{(\sigma_ {x_m}^{(I)})^2}}$. \footnote{This occurs in quantum optics when the cavity field is off high intensity, $\hat{x} \ket{\alpha} = \frac{a + a^\dag}{\sqrt{2}} \ket{\alpha} \approx  \frac{\alpha + \alpha^*}{\sqrt{2}} \ket{\alpha} = \text{Re}(\alpha) \ket{\alpha} = x \ket{\alpha}$} We will further assume that each path is not deformed by the gravitational interaction, which is reasonable given the small acceleration experienced by the masses~\cite{bose_spin_2017}. 
Thus the wavefunction of the combined $A$ and $B$ system at a time $\tau$ is given by:
\begin{align*}
    \ket{\Psi_{AB}(t_s + \tau)} &= \sum_{m,n= -j}^{j} c_m^{(A)} c_{n}^{(B)} e^{ i \pi \frac{g^2}{\omega_M^2} (m^2 + n^2)} \\
&\;\;\;\;\;\;\;\;\;\;\;\;\;\;\;\;\;\;\;\;\;\;  e^{-i \phi_{m,n}}\Ket{m ; x_m^{(A)}} \Ket{n ; x_n^{(B)}} \, ,
\end{align*}
where the phases are given by:
\begin{equation*}
\label{phase_linear}
    \phi_{m,n}^{\text{L}} = \frac{G M_A M_B \tau}{\hbar}\left| \Delta s + \Delta x (2 j + m - n) \right|^{-1} \, , 
\end{equation*}
\begin{equation*}
\label{phase_parallel}
    \phi_{m,n}^{\text{P}} = \frac{G M_A M_B \tau}{\hbar}\left| \Delta s^2 + \Delta x^2 (m - n)^2 \right|^{-1} \, ,
\end{equation*}
for the linear (L) and the parallel (P) set-ups. We took $\Delta x = \Delta y$ in order to give a comparison between the two set-ups. 
After the recombining process, the final state of the system (at a final time $t_f = 2 t_s + \tau$) is given by:
\begin{align*}
    \ket{\Psi_{AB}(t_f)} &= \biggl[ \sum_{m,n= -j}^{j} c_m^{(A)} c_{n}^{(B)} e^{ i 2 \pi \frac{g^2}{\omega_M^2} (m^2 - n^2)} \\
&\;\;\;\;\;\;\;\;\;\;\;\;\;\;\;\;\;\;\;\;\;\;  e^{-i \phi_{m,n}}\Ket{m} \otimes  \ket{n} \biggr] \otimes \ket{x_0^A; x_0^B} \, .
\end{align*}
The spin degrees of freedom are independent of the position ones, as the two masses are recombined in the center of their respective trap. Noticing the phases $\phi_{m,n}$, the remaining spin state is not separable and is entangled\footnote{With the considered approximation, the proposed protocol can be expressed as resulting from an evolution defined by a spin Hamiltonian. For instance, for the linear set-up:
\begin{equation*}
\hat{H}_G^{\text{L}} = \frac{GM_AM_B}{\left| \Delta s + \Delta x (2 j + \hat{J}_{z}^{(A)} - \hat{J}_{z}^{(B)} )  \right|}.
\end{equation*}}. Such entanglement will not arise in the case of a classical c-numbered gravitational Hamiltonian~\cite{bose_mechanism_2022}. 
\vspace{-0.2cm}
\section{\label{sec:entropy} von Neumann entropy}
\vspace{-0.2cm}
As a first measure of the entanglement, we will use the von Neumann entropy:
\begin{equation*}
    S(\rho_A) \coloneqq - \text{Tr} \left[\rho_A \ln (\rho_A ) \right],
\end{equation*}
where $\rho_A $ is the reduced density matrix of sub-system $A$. 

We will consider an experimental set-up as proposed in Ref. ~\cite{bose_spin_2017}, with $\tau = 2$ s, $M_A \approx M_B \sim  10^{-14}$ kg, and $\Delta x \approx 250 \,\mu $m, with the addition of Casimir screening, which allows $\Delta s = 50 \,\mu $m. 
Different spins state families will be compered, and their parameters will be optimized seeking to find the Maximal Gravity-induced Entanglement Entropy (MGEE) that can be generated for a given spin family. For conciseness, some of the plots of the numerical analysis will be reported in Appendix~\ref{app:entanglement}.  

\begin{figure}[!h]
  \centering
  \subfloat[J = 2]{\includegraphics[width=0.24\textwidth]{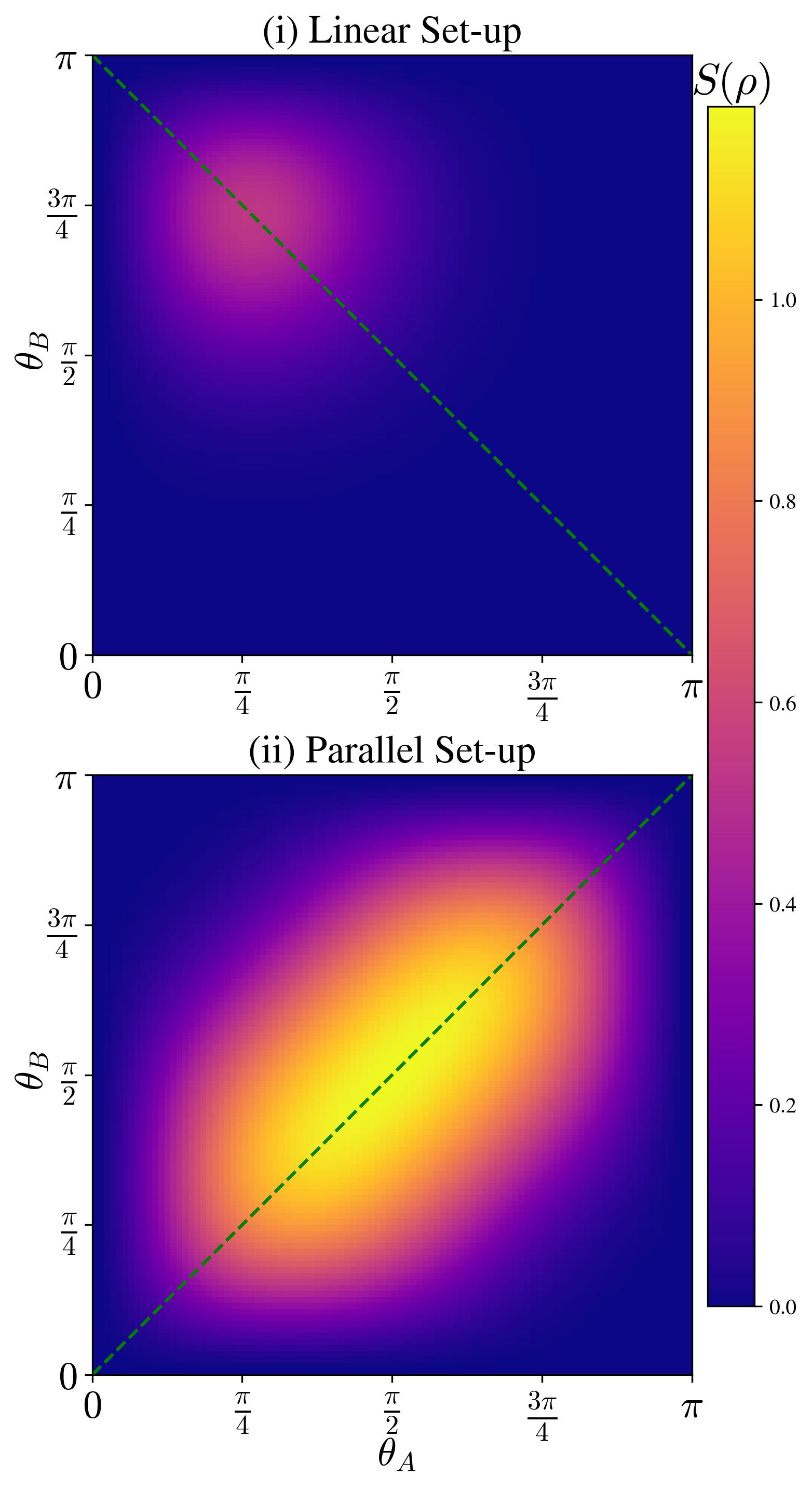}\label{plt:ent4}}
  \hfill
  \subfloat[J = 10]{\includegraphics[width=0.24\textwidth]{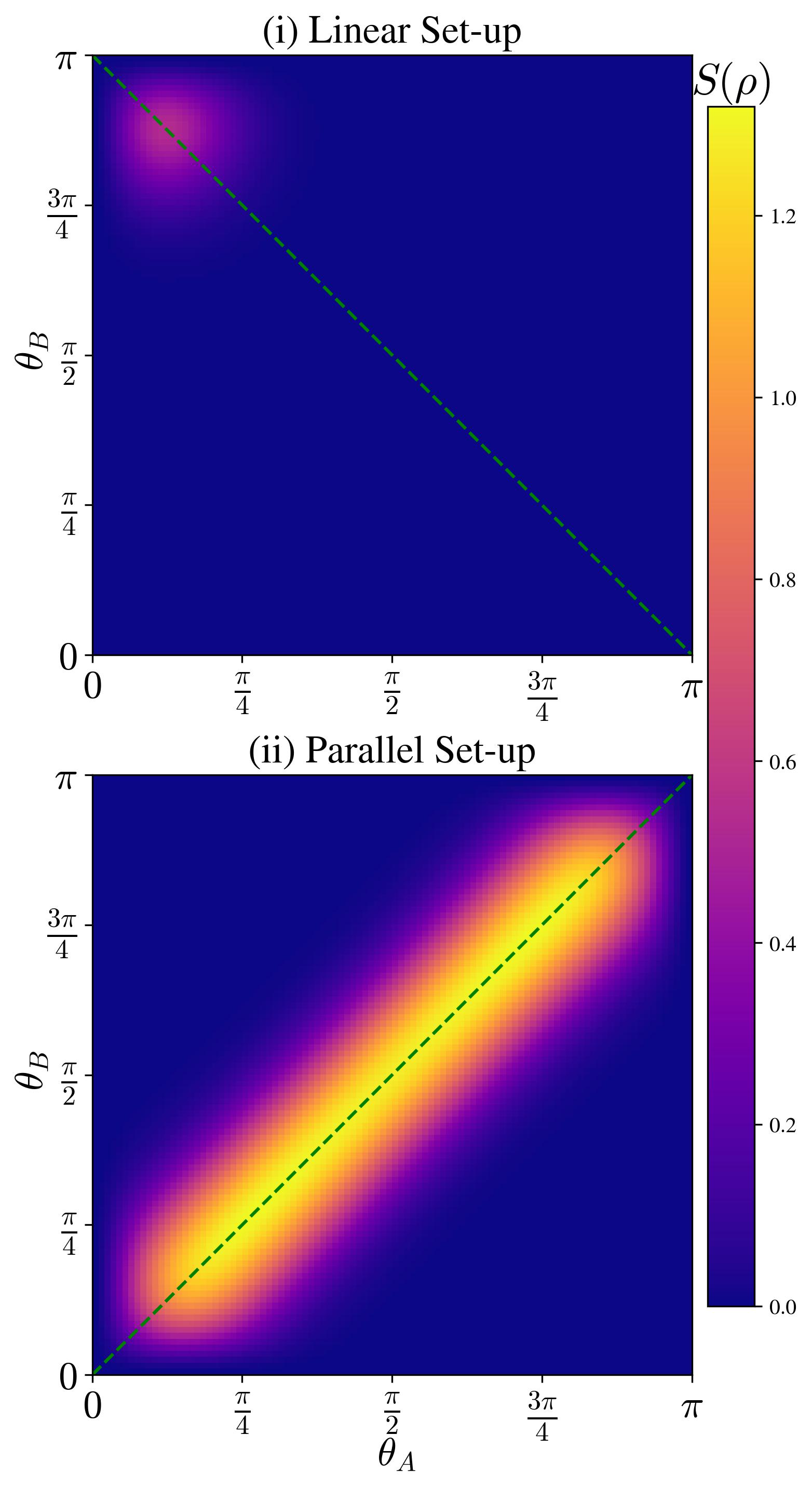}\label{plt:ent10}}
  \caption{\label{ent_thetas}Gravitational entanglement as a function of the rotation angles $\theta_I$ of two CSS with (a) $J=2$ and (b) $J=10$ in two SGS interferometers the (i) linear and (ii) parallel set-ups. The symmetry axes are in~green.\label{plt:ent_CSS}}
\end{figure}

Let us first study the case of CSS as \textit{easily} implementable in a laboratory setting. Each embedded CSS -- defined according to Eq.(\ref{SCS_state}) -- has two parameters $\phi_I$ and $\theta_I$, with $I \in \{ A, B\}$. Hence, finding the MGEE is an optimization problem with 4 parameters.
The von Neumann entropy has been found to be independent of the parameters $\phi_I$, and we will not further consider them. 
The gravitational entanglement as a function of $\theta_I$ is plotted in Fig.~\ref{ent_thetas} for both parallel and linear geometries. 

As expected, a significant entropy is produced when the mass probability distributions of the two interferometers are at their closest. 
In the linear set-up, this implies that most of the mass of the left interferometer is \textit{placed} on the right ($\theta_A$ small) while in the right interferometer the mass is on the left ($\theta_B$ large). In the parallel set-up, the mass probability distributions have to match on both sides, i.e. $\theta_A \approx \theta_B$. 
Higher spins require higher precision on the rotation angles $\theta_I$: for instance, in the parallel set-up, the requirement $\theta_A \approx \theta_B$ is more restrictive as $j$ increases as the entanglement region (yellow band in Fig.~\ref{plt:ent_CSS}) \textit{shrinks} in width. 

Hence, it is possible to maximize the entropy over $\theta_I$ and find the optimal rotation angles $\theta_I^{(o)}$ of the input CSSs for a given $j$ that gives the MGEE. 
The found optimal values respect the symmetries of the interferometers geometries: \mbox{ $\theta_A^{(o)} = \pi - \theta_B^{(o)}$ } for the linear set-up, and  $\theta_A^{(o)} = \theta_B^{(o)}$ in the parallel case. Thus, among the four parameters ($\theta_{A,B}, \phi_{A,B}$), only one is independent in the entanglement.

\begin{figure}[!h]
\centering
    \includegraphics[width=0.5\textwidth]{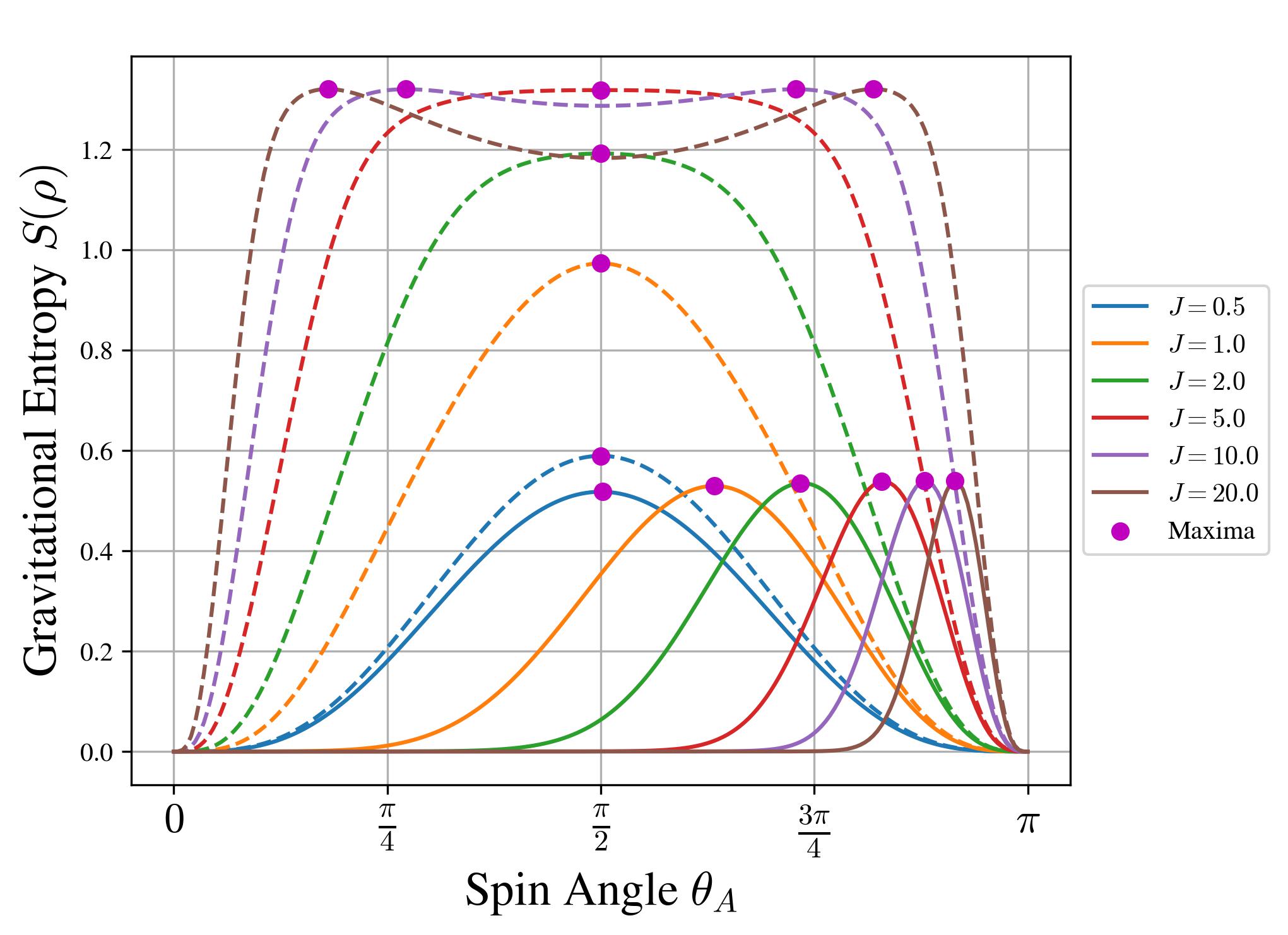}
     \caption{\label{ent_theta}Entanglement entropy as a function of the only independent parameter $\theta_A$,  for different spins $j$. The parallel set-up is plotted with dashed lines and the liner set-up with solid lines.}
\end{figure}

\begin{figure}[!b]
\centering
    \includegraphics[width=0.5\textwidth]{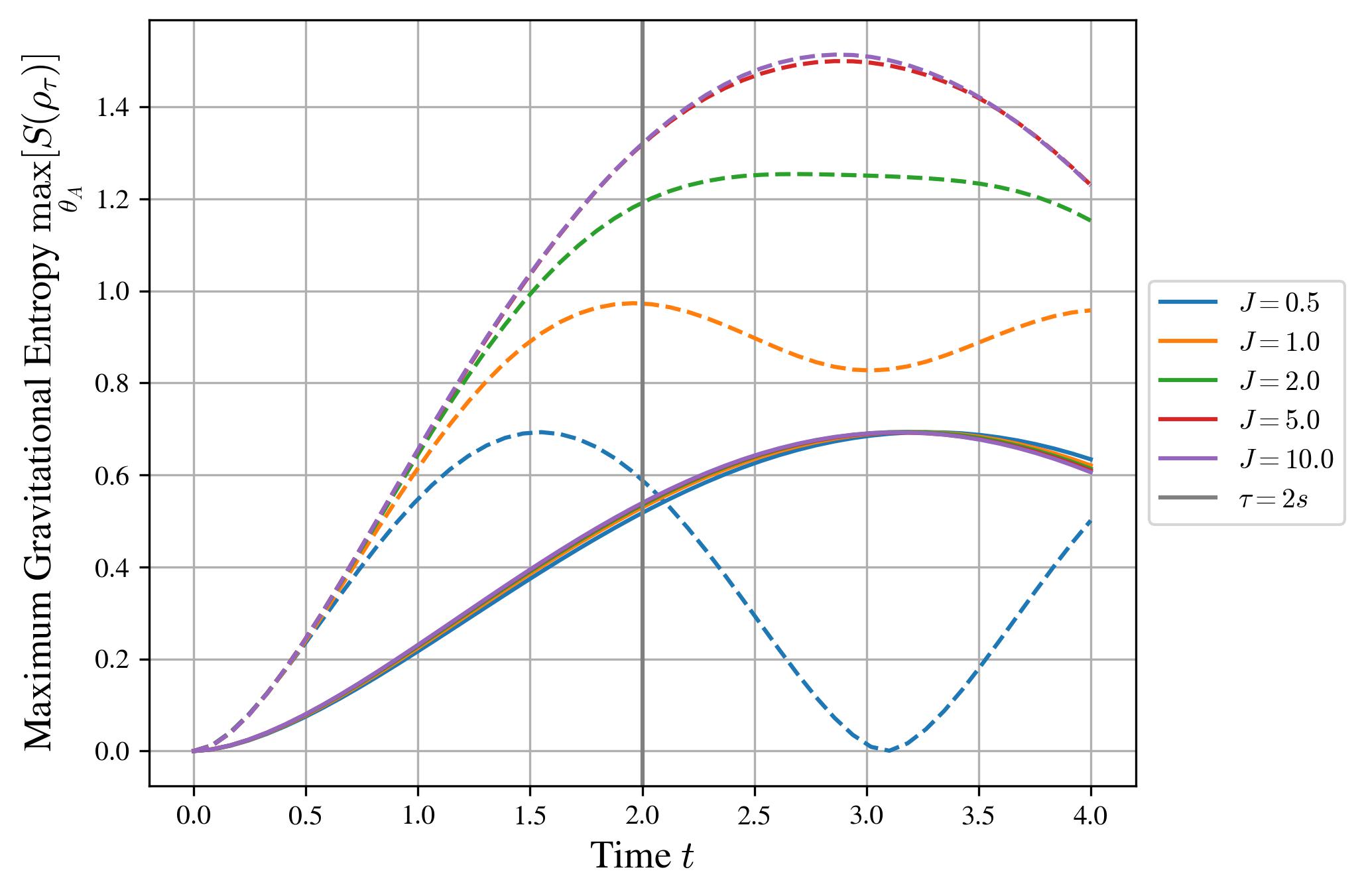}
    \caption{\label{ent_time} Time evolution of the MGEE generated by optimal CSS, $\ket{\theta_A^{(o)}}$. The parallel set-up is plotted with dashed lines and the liner set-up with solid lines.}
\end{figure}

The gravitational entanglement as a function of $\theta_A$ is plotted in Fig.~\ref{ent_theta} for different spins values and fixed time ($\tau=2$ s). 
In the linear set–up, $\theta_A^{(o)}$ increases with the spin value (and the mass is progressively concentrate on one side of the intefermoeter), while the MGEE remains approximately constant. In the parallel case and for small spins values ($j \le 5$), MGEE is at \mbox{$\theta_A = \pi/2$}. As the spin increases ($j > 5$), a local minimum is formed at $\theta_A = \pi/2$ and two maxima emerge symmetric with respect to  $\theta_A = \pi/2$. 
The time evolution of the MGEE can be computed for the optimal CSSs $\ket{\theta_A^{(o)} (\tau)}$, where $\theta_A^{(o)}$ becomes a time dependent parameter. This is plotted in Fig.~\ref{ent_time}.
Note that the entanglement is significantly enhanced in the parallel set-up. 
However, in the linear set-up, the time evolution of the optimal case does not differ significantly among different spins. 
This is also clear from Fig.~\ref{entropy_J} in Appendix~\ref{app:entanglement}, where the MGEE as a function of the spin value $j$ is plotted.

In the linear case, only a few $m$-components of the quantum state (the one closest to $-j$ for the left interferometer and to $+j$ for the right interferometer) generate entanglement. Given that there is not a significant improvement in increasing the Hilbert space in the linear set-up, following, we will only consider the parallel interferometric geometry. 

In the parallel set-up, the MGEE monotonically increases over $j$ (see Fig.~\ref{entropy_J}), tending to an asymptotic value. We will approximate the asymptotic value to be the MGEE for $j=30$ (as it is the largest spin computed). At short interaction time, the increment in the entanglement is less prominent for higher $j$, i.e. the MGEE approaches its asymptotic value at smaller $j$ for shorter interaction times $\tau$. Hence, to exploit the improvements of large spins, we require a large enough interaction time, for instance, as previously proposed, $\tau = 2s$.

Let us now consider a superposition of CSS, restricting the discussion to states of the form: 
\begin{equation}
\label{eq:supp}
\ket{\psi} = \mathcal{N} \left( \Ket{\frac{\pi}{2} + \delta \theta, \delta \phi} + \Ket{\frac{\pi}{2} -\delta \theta, - \delta \phi} \right) \, ,
\end{equation} 
with $\delta \theta, \delta \phi \in [0, \pi]$.\footnote{The case of $
\ket{\psi} = \mathcal{N} \left( \ket{\theta_I^{(o)} + \delta \theta, \delta \phi} + \ket{\theta_I^{(o)} -\delta \theta, - \delta \phi} \right) 
$ has also been studied, but no enhancement in the entanglement entropy were found for the cases of $\theta_I^{(o)} \neq\pi/2$, i.e. for $j>5$.} 

\begin{figure}[h]
    \includegraphics[width=0.49\textwidth]{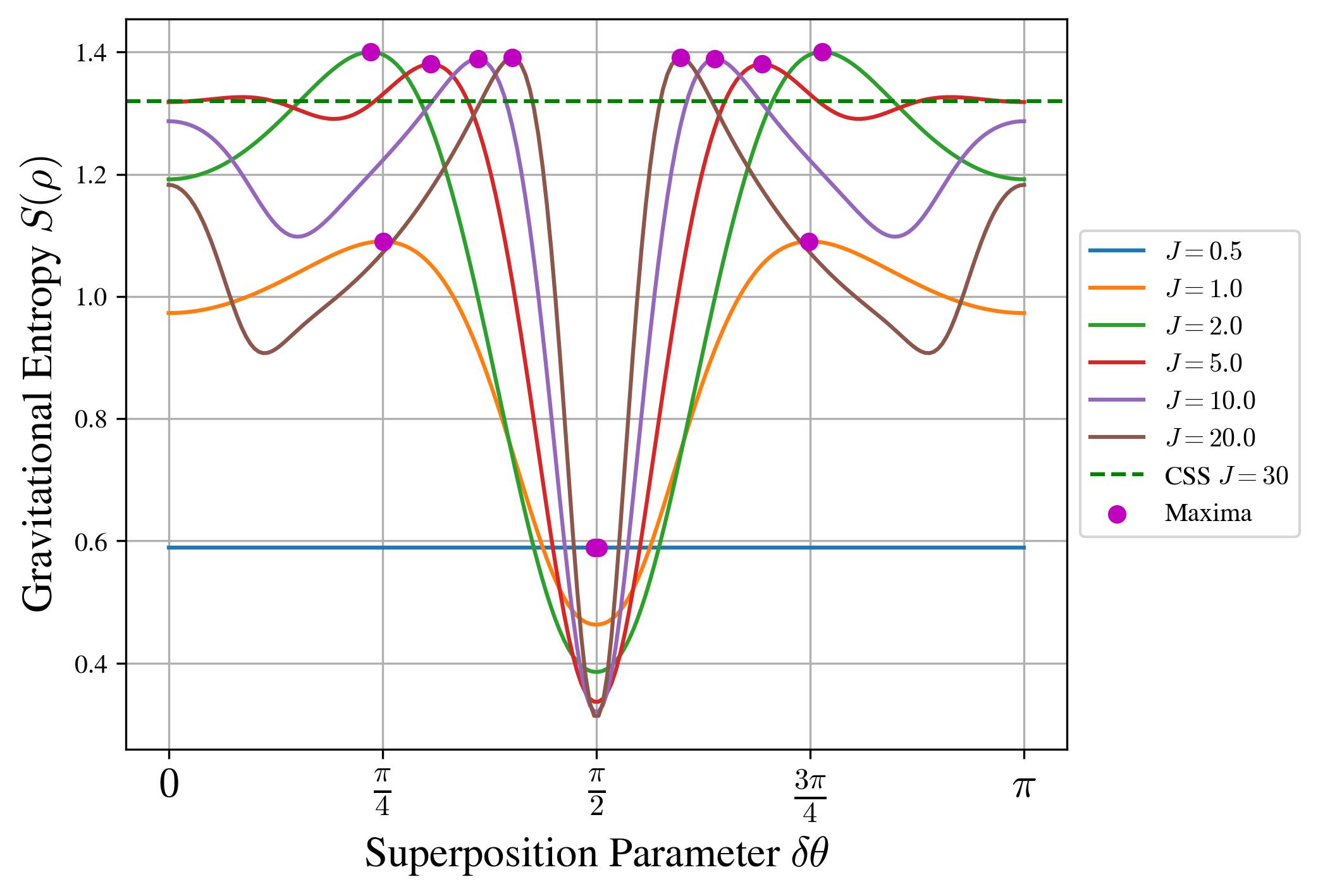}
        \caption{\label{plt:super_1} Gravitational entanglement at time $\tau = 2s$ between two superpositions of CSS (Eq.(\ref{eq:supp})) as a function of superposition parameter $\delta \theta$, describing the deviation from an initial CSS with $\theta = \pi/2$.} 
\end{figure}

The entanglement entropy as a function of $\delta \theta$ and $\delta \phi$ is plotted in Fig.~\ref{plt:super} in Appendix~\ref{app:entanglement}. The $\delta \phi$ variable leads to interference-like patterns, and this superposition parameter will be discarded as it does not significantly increase the entanglement entropy. There is an interesting dependency on the $\delta \theta$ variable, and the entanglement entropy as a function of $\delta \theta$ is plotted in Fig.~\ref{plt:super_1}. 

With a superposition of CSS, the MESS is higher with respect to the corresponding MGEE given by input CSS, for any given spin (with $j>1/2$, as the case of $j=1/2$ is trivial). Interestingly, for $j\geq2$, the MGEE for superpositions of CSS is larger than the MGEE asymptotic value for CSS. However, the MGEE is approximately constant for any $j \geq 2$. Hence, by creating a superposition of CSS, the entropy can be maximized for $j = 2$, and further increment of the spin value does not lead to significant improvements. The optimal superposition parameter, which maximizes the entropy, tends to $\pi/2$ as $j$ increases.\footnote{ A notable behaviour is the minima at $\delta \phi = \pi/2$, as it represents a NOON state case. Contrary to quantum sensing where such states are the most sensible, since they maximize the Fisher information, for gravitational entanglement detection, they do not represent a good choice of states, since they minimize the entanglement. }
Finally, it is possible to study SSS of the general form given in Eq.~(\ref{SSS_state}). The entanglement entropy can be computed as a function of the squeezing parameter ($\chi$), for both one-axis twisted and two-axis twisted states and of the input CSS angle $\theta$ (the spin state before the squeezing). This is plotted in Fig.~\ref{plt:squeezed_2} and \ref{plt:squeezed_1} in Appendix~\ref{app:entanglement}. It is found that using SSS does not significantly increase the MGEE. The behaviour becomes chaotic for higher squeezing parameters, where the SSS have higher entanglement entropy with respect to the corresponding CSSs. Such high squeezing is not reasonable in an experimental setting and so SSS does not represents a good input spin state. Hence, we will discard such states from our further analysis.

To summarise the main results, the MGEE for different families of spin state and spin values is given in Table~\ref{tbl:ent}. 
It seems that in terms of generating gravitationally induced entanglement for these experimental parameters, a superposition of coherent spin states with $j=2$ is most favorable. 

\begin{table}[!t]
\begin{center}
\begin{tabular}{ |p{ 0.14\textwidth}||p{ 0.07 \textwidth}|p{ 0.07 \textwidth}|p{ 0.07\textwidth}|p{ 0.07\textwidth}| }
 \hline
 \multicolumn{5}{|c|}{Maximal Gravity-induced Entanglement Entropy} \\
 \hline
Spin & $ j=1/2 $ & $j = 2$ & $j=5 $& $ j=10 $\\
 \hline
 \hline
 Coherent   & 0.59    & 1.19 &   1.32 &  1.32\\
 \hline
 Superposition & 0.59   & 1.40   & 1.38 & 1.39\\
 \hline
 1-axis Squeezed & 0.59  & 1.37 & 1.38 & 1.39 \\
 \hline
 2-axis Squeezed   & 0.59  & 1.22 & 1.36 & 1.33\\
 \hline
\end{tabular}
\end{center}
 \caption{\label{tbl:ent} Maximal Gravity-induced Entanglement Entropy generated by optimizing families of spin states in two GSG interferometer (parallel set-up) interacting via gravity with $\tau = 2$ s, $M_A \approx M_B \sim  10^{-14}$ kg, $\Delta x \approx 250 \,\mu $m, and $\Delta s = 50 \,\mu $m.}
\end{table}

\section{\label{sec:witness}Negativity}
The von Neumann entropy requires a full knowledge of the quantum state of the system and does not represent a suitable experimental observable. In particular, when describing the test masses as an open quantum system interacting with the environment (i.e. including decoherence), the system will entangle with the environment, and the entanglement entropy generated by gravity will be corrupted. Hence, the von Neumann entropy is not a good entanglement measure for globally mixed states.
In order to solve this issue, the negativity of the quantum state is considered~\cite{yczkowski_volume_1998, plenio_logarithmic_2005}. 
Similar to the PPT criterion\footnote{It has to be noted that in our analysis, the PPT criterion (which only consider the smallest negative eigenvalue) does not represent a good measure of entanglement, even if in principle it can detect entanglement. As $j$ increases, the smallest eigenvalue decreases (so the PPT criterion), while the number of negative eigenvalues increases. 
With the inclusion of decoherence such a quantity is then not robust, as it is too small.}, the negativity is a sufficient (but not necessary) measure of entanglement. 
The negativity is defined as the sum of the negative eigenvalues of the partial transposed (denoted with the superscript $PT$) density matrix of the system, i.e. $\braket{\mathcal{W}}  = \sum_{\lambda^{PT} < 0} \lambda^{PT} $. Note that -- according to our definition -- such a quantity is negative (while in other definitions, its absolute value is considered). It is possible to construct a witness operator that measures this quantity as
\begin{equation*}
    \hat{\mathcal{W}}=\sum_{\lambda < 0} \ket{\lambda} \bra{\lambda}^{PT},
\end{equation*}
where $\ket{\lambda} $ is the eigenvector corresponding to the $\lambda$ eigenvalue, and $PT$ represents the partial transposition of one of the Hilbert spaces. 

It is then possible to expand the $\hat{\mathcal{W}}$-operator in terms of a complete basis of observables which can be implemented in a laboratory setting. 
In the case of a spin $j$, these observables are the $2j+1$ dimensional irreducible representation of $SU(2)$, i.e. the generalized Gell-Mann matrices. 
The measurements expansion and statistics in terms of Gell-Mann matrices have already been studied for a \textit{qudit}, and we refer to the Appendix of Ref ~\cite{tilly_qudits_2021} for further explanation. 

As in the previous section, the negativity of the presented protocol for different spin states, spin values, and interaction time has been computed, to find the optimal spin state which minimize the negativity (recall that the negativity is a negative quantity). The behaviour of the negativity resembles the one of the entanglement entropy. For this reason, the plots are presented in Appendix~\ref{app:negativity}, while a summary of the results is given in Table \ref{tbl:nega}.
\begin{table}[!t]
\begin{center}
\begin{tabular}{ |p{ 0.14\textwidth}||p{ 0.07 \textwidth}|p{ 0.07 \textwidth}|p{ 0.07\textwidth}|p{ 0.07\textwidth}| }
 \hline
 \multicolumn{5}{|c|}{Minimal Gravity-induced Negativity} \\
 \hline
Spin & $ j=1/2 $ & $j = 2$ & $j=5 $& $ j=10 $\\
 \hline
 \hline
 Coherent   & -0.44   & -1.40 &   -2.02 &  -2.43 \\
 \hline
 Superposition & -0.44  & -1.76   & -2.61 & -3.02 \\
  \hline
\end{tabular}
\end{center}
 \caption{\label{tbl:nega} Minimal Negativity generated by optimizing families of spin states interacting via gravity.} 
\end{table}

Some qualitative differences between the von Neumann entropy and the negativity can be found from the plot in Appendix~\ref{app:negativity}. First of all, the dependency on the spin value of the minimal negativity is more prominent. 
The negativity monotonically decreases as function of $j$ toward an asymptotic value, which saturates at higher spin values. 
Hence, when considering this observable, the employment of larger spins become more preferable. Secondly, there is no local maximum in negativity arising for the rotation $\theta_{A}=\pi/2$ when the spin increases in value, and the optimal parameter \mbox{$\theta_{A}^{(o)}=\pi/2$} for any $j$ computed(in contrast with the entanglement entropy, where for $j \geq 5$, a local minima appeared). Thirdly, when considering superposition of CSS, the optimal $\delta \theta$ does not tend to $\pi/2$, but decreases in value.

\section{\label{sec:decoherence}Decoherence}

There are many sources of decoherence. The main sources of decoherence were already mentioned in~\cite{bose_spin_2017}. Then, the analysis of~\cite{van_de_kamp_quantum_2020, pedernales_motional_2020,rijavec_decoherence_2021,tilly_qudits_2021,schut_improving_2022} have reinforced the constraints along with external jitters and gravity gradient noise~\cite{toros_relative_2021}, gravitational decoherence~\cite{toros_loss_2020,wu_quantum_2021}, dipole-induced decoherence~\cite{fragolino_decoherence_2023}, decoherence/dephasing due to the plate~\cite{schut_relaxation_2023}, and gravity-induced decoherence due to the apparatus~\cite{gunnink_gravitational_2023}. These computations are based on earlier analysis of decoherence in a quantum system~\cite{Romero_Isart_2011, chang_cavity_2009}. 
There are also sources of decoherence due to internal phonon vibration~\cite{henkel_internal_2022,henkel_universal_2023}, and due to the effect of rotation~\cite{japha_quantum_2023,ma_torque_2021}.

The aim of this section is to give a first comparative analysis of decoherence between different spins. Overall, sources of decoherence can be divided into two groups: \textit{external} decoherence (due to scattering of external particles) and \textit{internal} decoherence (due to internal structure of the mass). We will consider the former but not the latter, as we assume the internal structure of the mass to be independent of the spin. However, as we will see, larger spins increase the total superposition and the Hilbert space, hence changing the behaviour of the system under some sources of external decoherence.

Following Ref. ~\cite{schlosshauer_quantum_2007}, the decoherence due to scattering of external particles with de Broglie wavelength $\lambda_B$ onto a mass in a superposition $\Delta x$ can be further divided into two subgroups, namely, \textit{short-wavelength} and \textit{long-wavelength} limits. In the case of the former ($\lambda_B \ll \Delta x$), a single scattering event would resolve all the \textit{which-path} information of the superstition.\footnote{For instance, at T = 3 K, the de Broglie wavelength of $\text{O}_2$ air molecules is $ \lambda_\text{dB} \approx 2 \cdot 10^{-10} $m which is much smaller compared to $\Delta x$ considered in the experiment.} While in the case of the latter ($\lambda_B \gg \Delta x$), multiple scattering events are needed to resolve the information.\footnote{For black-body photons and $T\sim 1$ K, \mbox{$\lambda_\text{BB} \sim 10^{-3}\gg \Delta x$}.}

Both long and short wavelength decoherence limits are compared for different spins in regard to how the negativity is affected. To give a meaningful comparison, the decoherence rates are kept as a variable (as they are dependent on the specifics of the experiment - material of the mass, temperature and pressure). The specifics on how scattering decoherence is included in the model is given in Appendix~\ref{app:dechoerecce}. 
Fig.~\ref{plt:nega_dec_two} plots the Negativity as a function of the rotation angle $\theta_A$ of the input CSS for different decoherence rates.

\begin{figure}[!tbp]
  \centering
  \subfloat[Short-wavelength]{\includegraphics[width=0.24\textwidth]{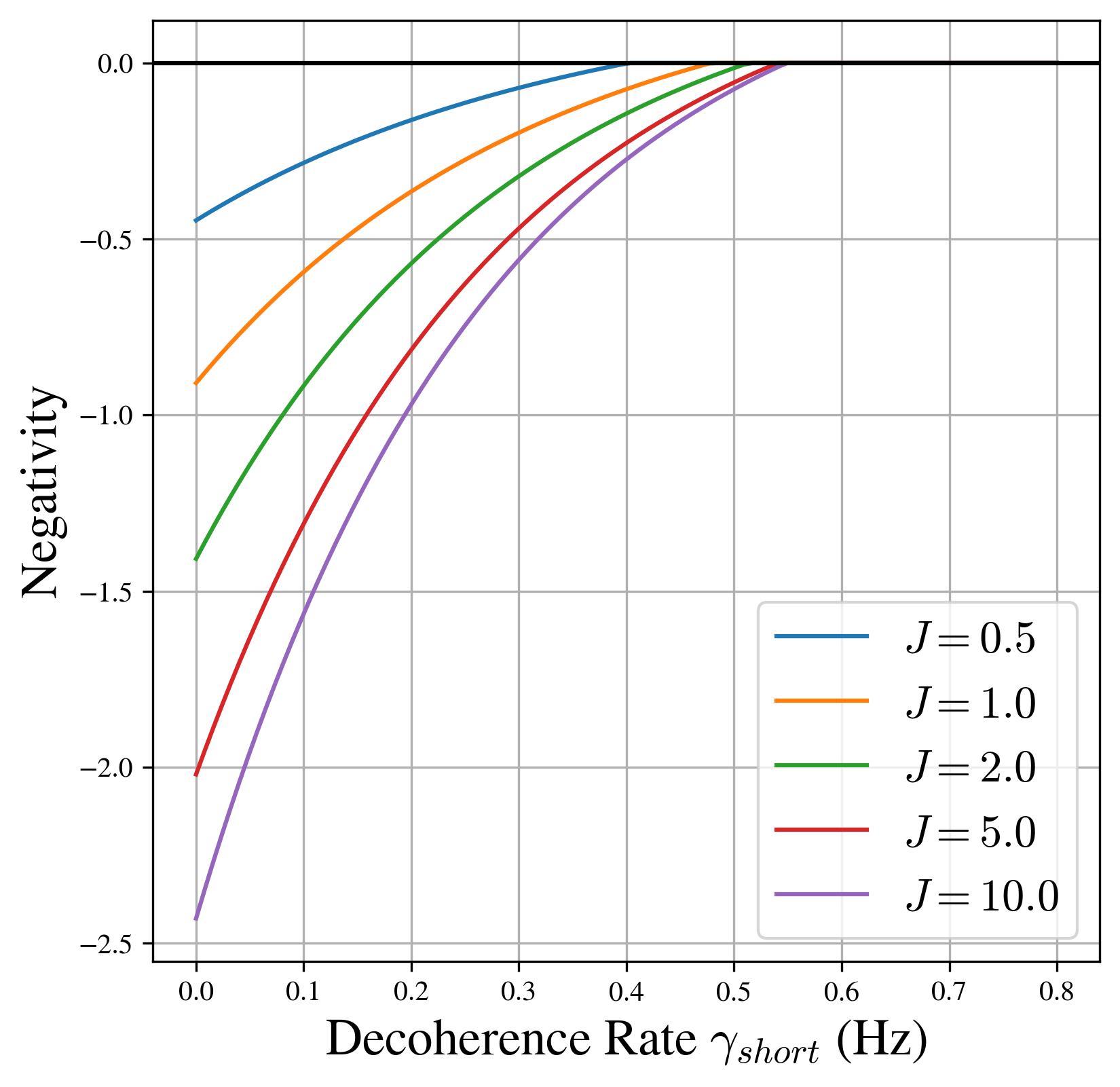}}
  \hfill
  \subfloat[Long-wavelength]{\includegraphics[width=0.24\textwidth]{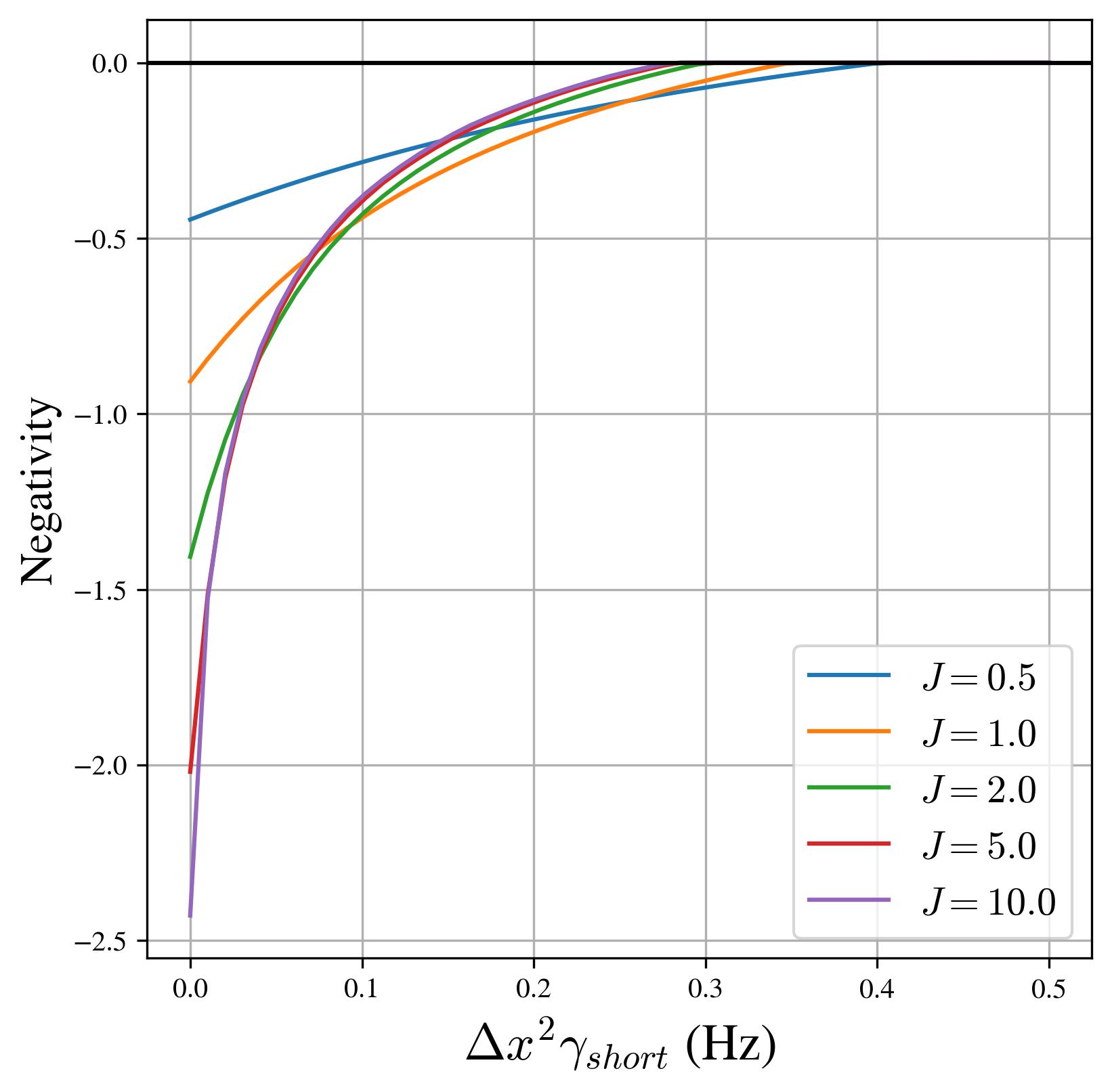}}
  \caption{\label{plt:nega_dec} Minimal negativity for different spins values as a function of the decoherence rate for (a) short-wavelength and (b) long-wavelength limits.}
\end{figure}

In the case of short-wavelength, the optimal angle is still $\theta_A^{(o)} = \pi/2$, for any decoherence rate $\gamma_{short}$. However, in the case of long-wavelength limit, the optimal angle changes according to the quantity $\Gamma_{long} \Delta x^2$.
\begin{figure}[!tbp]
  \centering
  \subfloat[Short-wavelength]{\includegraphics[width=0.24\textwidth]{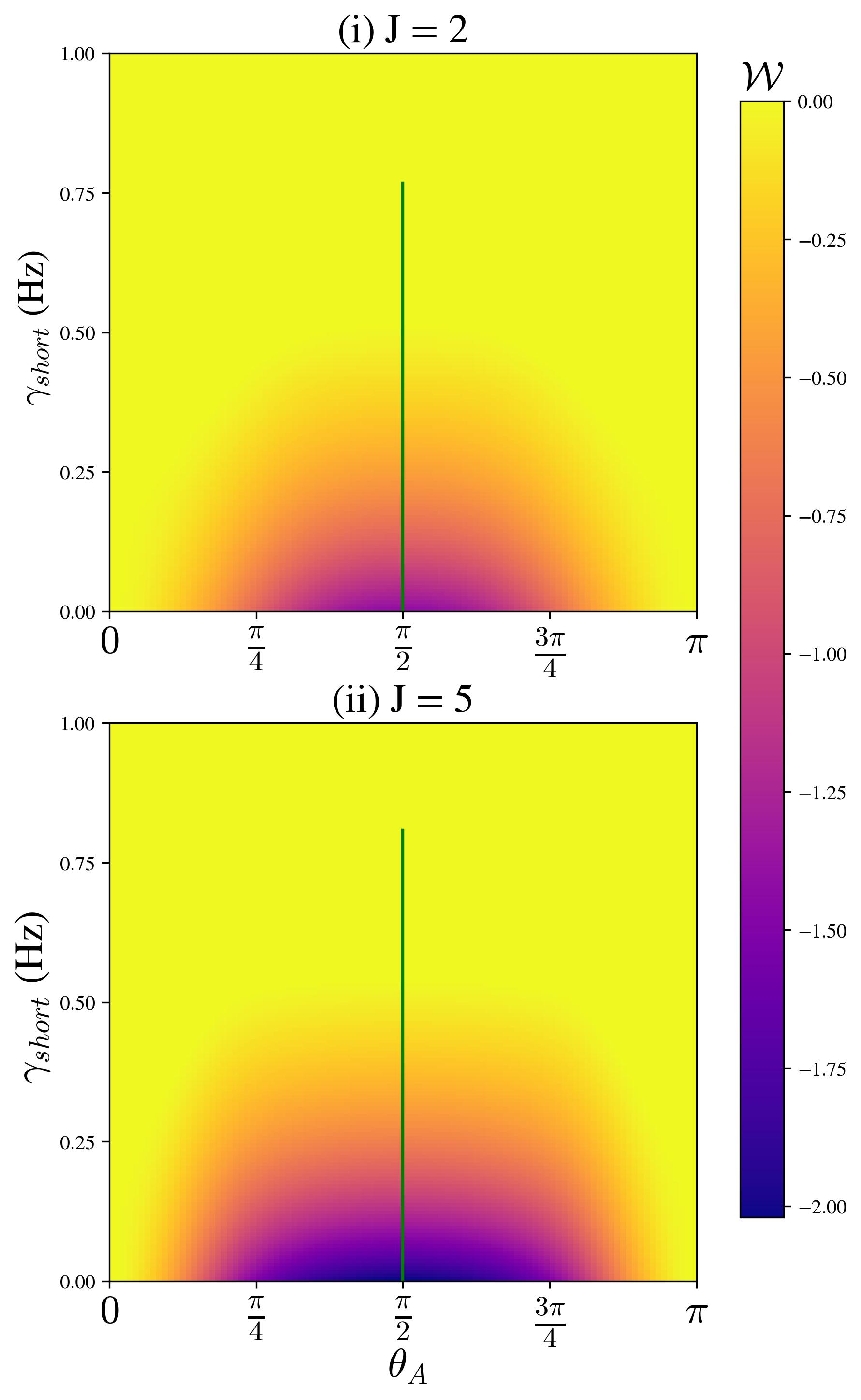}\label{plt:nega_short_rot}}
  \hfill
  \subfloat[Long-wavelength]{\includegraphics[width=0.24\textwidth]{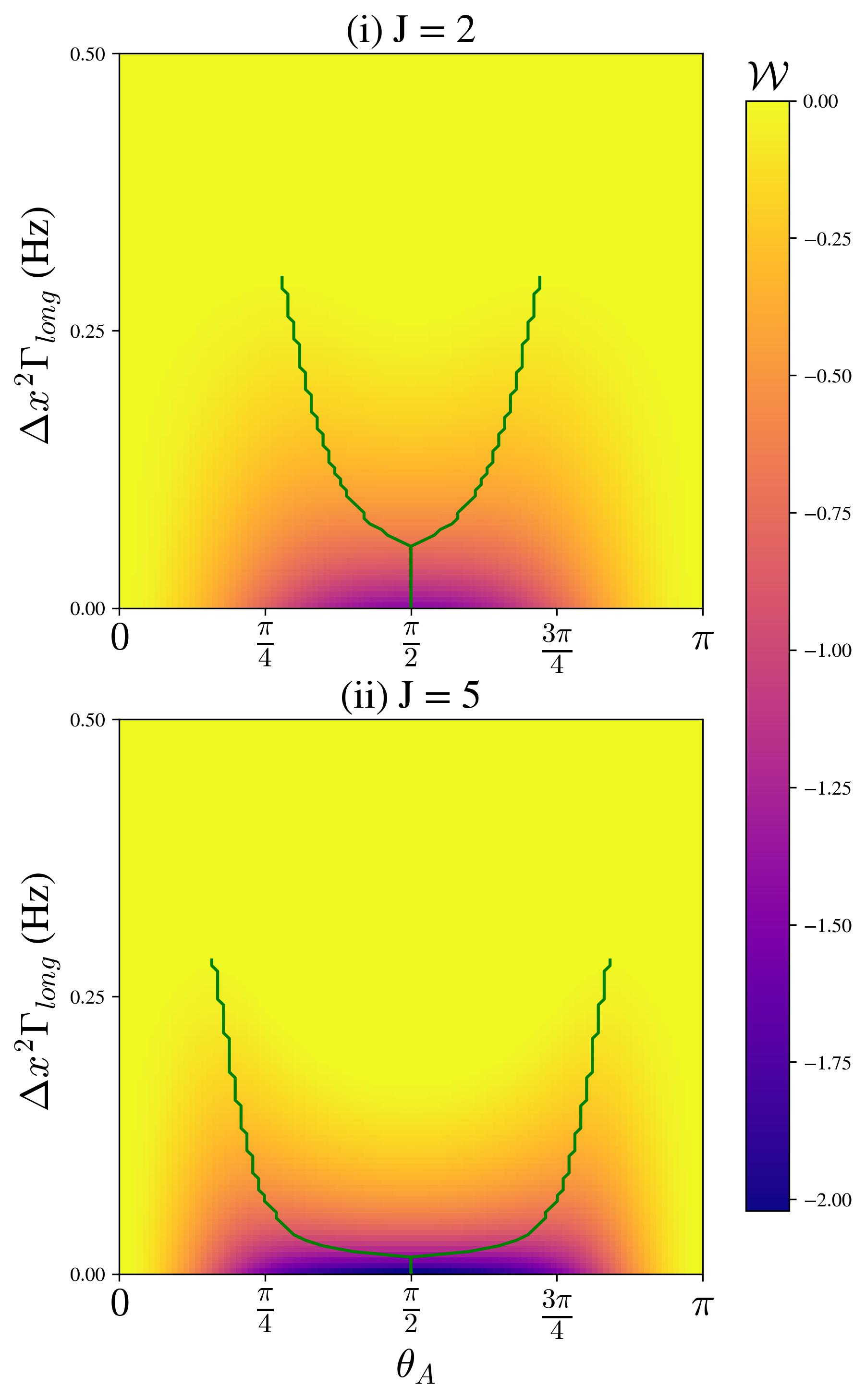}\label{plt:nega_long_rot}}
  \caption{\label{plt:nega_dec_two}Negativity as a function of the rotation angle and decoherence rate for (a) short-wavelength and (b) long-wavelength limits. The optimal angle as a function of the decoherence rate is plotted in green.}
\end{figure}
As was done previously, the negativity can be minimized over input spin states at different decoherence rates. 
The results on the minimal negativity generated by gravity with decoherence are plotted in Fig.~\ref{plt:nega_dec}.
In the short-wavelength limit, larger spins are more resilient under decoherence but, conversely, are more sensitive to long-wavelength decoherence sources. This is expected as larger spins imply a larger total superposition distance, which is linear in $j$. On one hand, the dependency on the superposition distance in the long-wavelength limit leads to a worse performances of large spins, which have a larger superposition. 
On the other hand, in the short-wavelength limit, a larger superposition does not affect the decoherence, as a scattering event would completely resolve the path taken even for the smallest superposition ($j=1/2$). In an experimental scenario, black-body noise (main source of long-wavelength decoherence) can be significantly reduced by cooling the system and the environment with temperature $T<1$ K, as the decoherence rate scales as $T^9$~\cite{Romero_Isart_2011,schlosshauer_quantum_2019}. However, for molecules scattering (short-wavelength), a high-vacuum is required, which may represent a bigger experimental challenge. 
With this in mind we conclude that large spins are more resilient under the \textit{most challenging} source of external decoherence, namely, air molecules scattering.

\section{\label{sec:conclusion}Conclusion}

An arbitrary spin embedded in a mesoscopic mass undergoes splitting in $2 j + 1$ paths when a magnetic gradient is applied, representing a generalization of the Stern-Gerlach interferometer to an arbitrary spin. The total spatial superposition size increases linearly in $j$ while the splitting time remains constant. Therefore, the presented a dynamics exemplifyies a controlled protocol to entangle any spin state with a spatial degree of freedom. This creates a joint state which includes a spatial superposition of coherent states, weighted accordingly to the spin distribution in the quantization axis representation. Spin coherent states, superposition of these, and squeezed states have been used in this study. 

The entanglement generated by two adjacent GSG interferometers interacting via gravity as quantified by the von Neumann entropy, witnesses the quantum nature of the gravitational field. 
The aforementioned spin state families were varied in order to find the optimal spin state which maximizes the gravitational entanglement, and, similarly, which minimize the negativity both with and without decoherence.

Such a procedure is first applied to the case of coherent spin states, which are the most easily implementable spin states in experiments. Two interferometer geometries are considered, namely, the linear and parallel set-ups. With the former, no improvements are found, as only a small fraction of the state is involved in the generation of the entanglement. In the latter, the maximum entanglement entropy monotonically increases as a function of the spin and reaches an asymptotic value for a given interaction time. A similar behaviour is found for the minimal negativity, which monotonically decreases with the size of the Hilbert space. For instance, by considering a spin $j=10$, the entanglement entropy is increased of a factor of $2.2$ with respect of the spin-$1/2$ case (and, equivalently, the negativity -- an entanglement monotone -- decreased of a factor of $5.5$).

Superposition of coherent spin states can be used to further increase the optimal entropy (negativity), which is increased of a factor of $2.5$ ($6.9$) for \mbox{$j=10$}. Squeezed spin states, with both one-axis and two-axis twisted Hamiltonian, do not significantly increase the generated entanglement entropy. The experimental challenges of creating these last two families of spin states may overcome the small increment in the entanglement entropy, compared to the CSS. 


The presented dynamics are of particular interest when considering NV-center nano-diamonds. It is a known experimental challenge to create a single-center nano-diamond, as multiple vacancies may be present. Until now, the dynamics of a multiple NV-centers mass in a magnetic gradient was unknown. 
With the assumption that the coupling with the magnetic gradient is the same among every center and neglecting the spin-spin coupling, the mass will evolve according to the proposed GSG interferometer.

The challenges of creating a mesoscopic superposition in an experimental setting are numerous and further studies should be carried out. 
The diamagnetic response from a mass in a magnetic gradient behaves as a trap (which allowed the presented interferometer) but also significantly increases the splitting time. 
As such a phenomenon is present both in this study, as well as in the original protocol, we have, in a sense, compared them here in an equal footing. 
Further works on low decoherence environments are encouraged in order to create such a large and massive superposition. 
An alternative solution to this issue may be achieved from a material science prospective with the design of a low diamagnetic susceptibility mesoscopic mass.  

To conclude, it is of great importance to experimentally study quantum aspects of the gravitational field. The experimental challenge of reaching mesoscopic masses in a quantum regime demands theoretical developments to find protocols that are more sensible to quantum phenomena of gravity. The question that we must pose is: \textit{Does the employment of large spins -- i.e. a $2j + 1$ spatial superposition of a mass weighted according to a spin distribution -- give a significant improvement to make quantum natured gravity-induced entanglement detection achievable in an experiment?} While improvements in the entanglement and a witness were found, we conclude that the entanglement cannot be increased of the orders of magnitude required to conduct the experiment with the current quantum technologies. However, once significant developments regarding noise isolation and splitting dynamics are achieved, the present protocol may lead to the necessary increment which may distinguish detectable and undetectable entanglement signals.

\section*{Acknowledgements}
L.B. would like to aknowledge Engineering and Physical Sciences Research Council (EPSRC) grants (EP/R513143/1 and EP/W524335/1), supported, in part, by Perimeter Institute for Theoretical Physics (Research at Perimeter Institute is supported by the Government of Canada through the Department of Innovation, Science and Economics Development and by the Province of Ontario through the Ministry of Colleges and University). 
S.B. would like to acknowledge EPSRC grants (EP/N031105/1, EP/S000267/1, and EP/X009467/1) and grant ST/W006227/1. We thank the Center for Information Technology of the University of Groningen for their support and for providing access to the H\'abr\'ok high performance computing cluster.


%

\clearpage

\appendix 

\onecolumngrid

\section{Diamagnetic Hamiltonian}\label{app:diamagnetic}

A uniform magnetic field applied to a diamagnetic material
leads to a diamagnetic response. 
For a suitable choice of magnetic profile, this can be treated as a quantum harmonic oscillator trapping the mass, where the frequency of the magnetic field is related to the frequency of the harmonic oscillator.
Additionally, the spin embedded in the diamagnetic material will couple to a magnetic field gradient in a Stern-Gerlach-like interaction.
For a general magnetic field $\bf{B}$, the Hamiltonian of the diamagnetic system reads: 
\begin{equation*}
    \hat{H} = \frac{\hat{\bf{p}}^2}{2 M} - \frac{M}{2} \frac{\chi_m }{\mu_0} \hat{\bf{B}}^2 + \tilde{g} \mu_B \hat{\bf{J}} \cdot \hat{\bf{B}} \, ,
\end{equation*}
where $\hat{\bf{p}}$ and $ \hat{\bf{J}} $ are the 3-dimensional momentum and spin operators of the mass M, respectively, $ \hat{\bf{B}} $ is the magnetic field operator, $\chi_m$ the mass (density) magnetic susceptibility, $\mu_0$ the vacuum permeability,  $\mu_B$ is the Bohr magneton and $\tilde{g}$ is the Landé g-factor.

By choosing the magnetic gradient of the form $\bm{B} = \left(\partial_x B \right) x \bm{e}_z $ (following~\cite{pedernales_motional_2020}), where $\partial_x B $ is the strength of the magnetic gradient and $\bm{e}_z$ is the unit vector align with the $z$ spin axis, the Hamiltonian on the $x$-axis (the others axes have free evolution) can be further simplifies to:
\begin{align*}
    \hat{H}_x = \frac{\hat{p}_x^2}{2 M} - \frac{1}{2} M \frac{\chi_m \left(\partial_x B \right)^2}{\mu_0}  \hat{x}^2 +\tilde{g} \mu_B \left(\partial_x B \right) \hat{J}_z \hat{x} \, ,
\end{align*}
where we restricted the Hamiltonian to the splitting axis, $\hat{x}$. 

Treating the trapped systems as a quantum harmonic oscillator, we can determine the frequency of the oscillation. 
Consequently we determine the coupling constant $g$ used in Eq.~\ref{eq:coupling}, which was defined as the coefficient of the term $\hbar \hat{J}_z(\hat{a}+\hat{a}^\dagger)$ in the Hamiltonian, where the creation and annihilation operators come from the definition $\hat{x} = \sqrt{\hbar/(2m\omega)}(\hat{a}+\hat{a}^\dagger)$.
\begin{equation*}
    \omega_M = \sqrt{\frac{|\chi_m|}{\mu_0}} \left(\partial_x B \right) \;\;\;\; \implies   \;\;\;\;    g =  \tilde{g} \mu_B \sqrt{ \frac{\left(\partial_x B \right) }{2 \hbar M } \sqrt{\frac{ |\chi_m|}{\mu_0 } }}
\end{equation*} 
and the Hamiltonian becomes the one given in Eq.~(\ref{splitting_hamiltonian}). Finally, the superposition distance (see Eq.~\ref{eq:deltax_MS}) and splitting time can be computed:
\begin{equation*}
   t_s \coloneqq \frac{\pi}{\omega_M} = \frac{\pi}{\left(\partial_x B \right)} \sqrt{\frac{\mu_0}{|\chi_m|}} \, ,\;\;\;\;\;\;\;\;\;\; \Delta x = \frac{2 \tilde{g} \mu_B \mu_0}{M |\chi_M| \left(\partial_x B \right)}
\end{equation*}
Considering a diamond with $|\chi_M| \approx 6 \cdot 10^{-9} \,\mbox{m}^3/\mbox{kg}$ ~\cite{heremans_1994} and $M\approx 10^{-14}$ kg:
\begin{equation*}
   t_s \approx 44.7 \left| \left(\partial_x B \right) \right|^{-1} \mbox{s} \, ,\;\;\;\;\;\;\;\;\;\; \Delta x = 7.5 \cdot 10^{-7}  \left| \left(\partial_x B \right) \right|^{-1} \mbox{m} \, .
\end{equation*}
In order to reach the wanted superposition of $\Delta x \approx 2.50 \cdot 10^{-4} $m, a magnetic gradient $ \left(\partial_x B \right) \approx 3.0 \cdot 10^{-3} \text{T/m}$, which implies a splitting time of $ t_s \approx 1.4 \cdot 10^{4} $s.
This is the well-known diamagnetic \textit{issue} (also present in the original proposal): a large superposition requires a weak magnetic gradient which significantly increases the splitting time.

\section{Splitting Unitary Evolution}
\label{app:unitary_evolution}
Using time in frequency unity, the Hamiltonian in Eq.~(\ref{splitting_hamiltonian}) gives the evolution:
\begin{equation}
\label{unitary_original}
U_I\left( t \right) =  e^{ -it  a^\dag a + it k J_z \left(a^\dag + a\right)}
\end{equation}
where $ r = \frac{\omega_0}{\omega_M}$ and $ k = \frac{g}{\omega_M}$. 
Using Schwinger’s Oscillator model of angular momentum, the spin algebra can be expressed in term of the algebra of two uncoupled quantum harmonic oscillators~\cite{sakurai_jun_john_modern_1994}. Each quantum harmonic oscillator is associated with ladder operators $b_+$ and $b_+^\dag$ ($b_-$ and $b_-^\dag$) creating and annihilating a particle with $m = \frac{1}{2}$ ($m = - \frac{1}{2}$). 
In this formalism, the projection of the angular momentum on the $z$-axis is equal to half the difference between of quanta in the $+$ state and in the $-$ state, $J_z = \frac{1}{2} \left( b_+  b_+^\dag - b_-  b_-^\dag  \right)$. 
Applying this to Eq.~(\ref{unitary_original}) recovers the unitary evolution studied in ~\cite{bose_preparation_1997}, with two commuting quantum harmonic oscillators instead of one. For completeness, we will derive the simplified unitary evolution. Following ~\cite{bose_preparation_1997}, let us define the \textit{spin-displacement} operator:
\begin{equation*}
    T = e^{ - \frac{k}{2} \left( b_+^\dag b_+ - b_-^\dag b_- \right) \left(a^\dag -a\right)} \, .
\end{equation*}
Using the Baker-Cambell-Hausdorf (BCH) expansion, the following identities can be derived:
\begin{align*}
T a T^\dag &= a +  \frac{k}{2}   \left( b_+^\dag b_+ - b_-^\dag b_- \right) \, ,\\
T a^\dag T^\dag &= a^\dag +  \frac{k}{2}  \left( b_+^\dag b_+ - b_-^\dag b_- \right) \, ,\\
T b_\pm^\dag  b_\pm T^\dag &= b_\pm^\dag  b_\pm \, .
\end{align*}
The last one implies that $T J_z T^\dag = J_z$. Recalling that $U f \left( \{ X_i\} \right) U^\dag =  f \left( \{ U  X_i  U^\dag \} \right) $, for any $U$ unitary:
\begin{equation*}
 T U_I\left( t \right)  T^\dag = e^{ i t k^2 J_z^2} e^{ -it  a^\dag a   } \, .
\end{equation*}
Multiplying on the left (right) by $T^\dag$ $(T)$, the unitary evolution can be rewritten as:
\begin{equation*}
U_I\left( t \right) =
e^{ it k^2 J_z^2 }  e^{- k J_z^\dag \left(a - a^\dag\right)} e^{-it  a^\dag a } e^{ - k J_z \left(a^\dag - a\right)} \, .
\end{equation*}
Furthermore, noticing that
\begin{equation*}
    e^{-ita^\dag a } \left( J_z (a^\dag - a) \right) e^{ita^\dag a } =  J_z   \left( a^\dag e^{ - it}  -  a e^{ it} \right) \, ,
\end{equation*}
the unitary can be further simplified to:
\begin{equation*}
U_I\left( t \right) = 
e^{ it k^2 J_z^2  }  e^{ - k J_z  \left(a - a^\dag\right)} e^{ - k J_z  \left( a^\dag e^{ - it}  -  a e^{ it} \right) } e^{-ita^\dag a } \, .
\end{equation*}
Finally, applying again the BCH expansion:
\begin{equation*}
U_I\left( t \right) = 
e^{ ik^2 J_z^2 \left(t - \sin(t) \right) }  e^{ k J_z   \left( \eta (t) a^\dag  - \eta (t)^* a  \right) } e^{-ita^\dag a} \, ,
\end{equation*}
where the function $\eta (t) = 1 - e^{-it}$ was defined. We note the $\eta (t) b^\dag  - \eta (t)^* b$ component of the unitary evolution, is typical of a time-dependent displacement operators in the position basis, $\mathcal{D} \left(\eta (t)\right) = e^{  \left( \eta (t) b^\dag  - \eta (t)^* b  \right) } $ . Thus, recalling that one of the definitions of position coherent state is the displacement of the vacuum, i.e. $\ket{\alpha} \coloneqq \mathcal{D}(\alpha) \ket{0} $, the quantum state in Eq.~(\ref{quantum_state_evolution}) follows trivially.

\section{\label{app:entanglement}Additional Entanglement Results}
In this Appendix, some of the plots regarding the vonNeumann entropy generated by gravity are presented. In Fig. \ref{entropy_J}, the maximized entanglement entropy for input CSSs is plotted as function of the total anglular momentum $j$. In Fig. \ref{plt:super}, the entanglement entropy is plotted as function of $\delta \theta$ and $\delta \phi$ for a superposition of CSSs defined according to Eq.~(\ref{eq:supp}). In Fig. \ref{plt:squeezed_2} and \ref{plt:squeezed_1}, the entanglement entropy for SSS is ploted.

\begin{samepage}
\begin{figure}[!h]
\centering
    \includegraphics[width=0.6\textwidth]{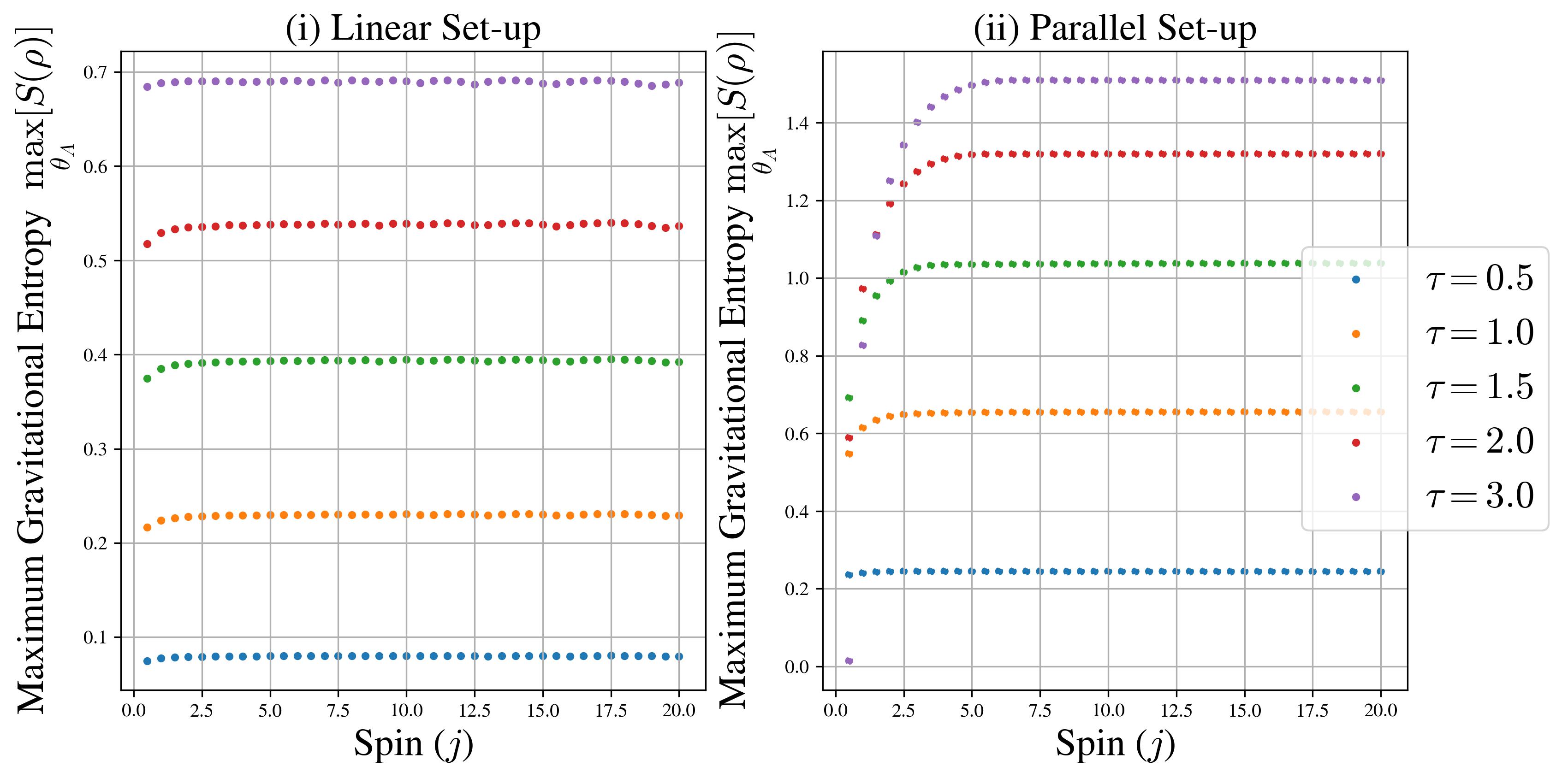}
    \caption{\label{entropy_J} Maximum entanglement entropy generated by optimal CSS $\ket{\theta_A^{(o)}}$ for (a) the liner and (b) the parallel set-ups, as a function of the spin $j$, at $\tau = 2s$}
\end{figure}

\begin{figure}[!h]
  \centering
  \subfloat[Superposition of CSS]{\includegraphics[width=0.35\textwidth]{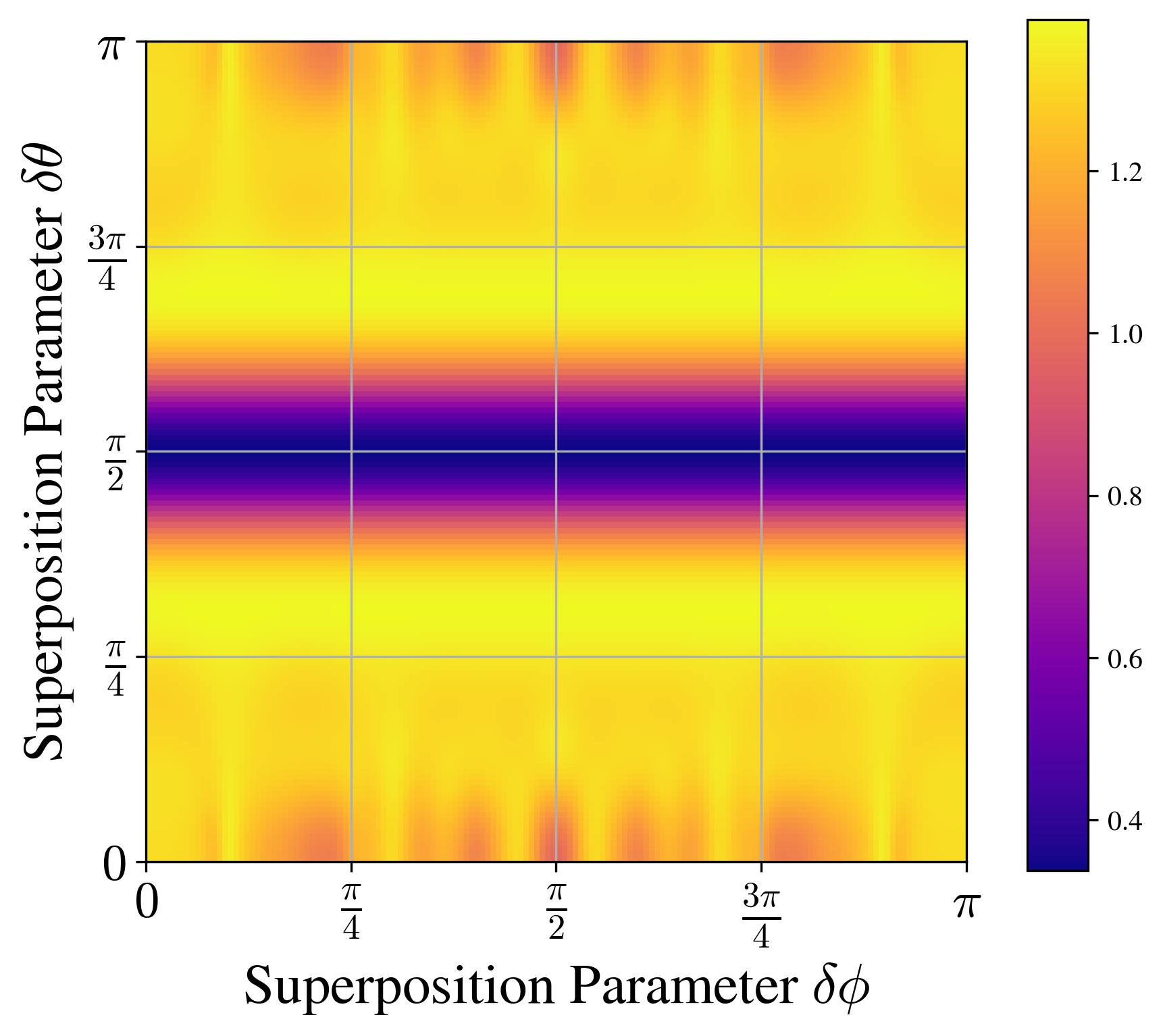}\label{plt:super}}
  \subfloat[Squeezed Spin State]{\includegraphics[width=0.55\textwidth]{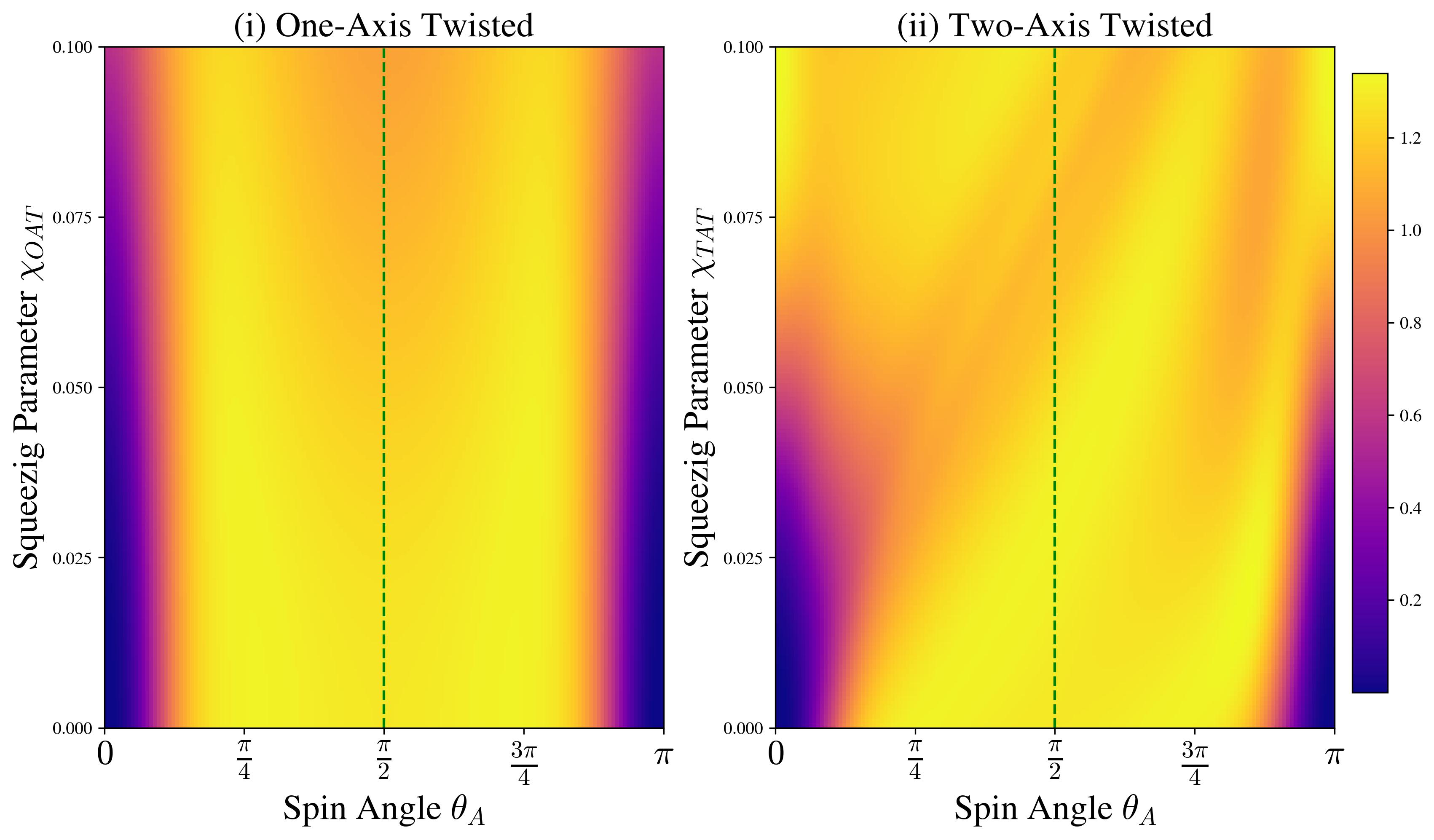}\label{plt:squeezed_2}}
  \caption{ Gravitational entanglement of (a) superpositions of CSS (Eq.~(\ref{eq:supp}), $J=5$) as a function of $\delta \theta$ and $\delta \phi$ with $J=5$ and (B) squeezed spin state ($J=10$) as a function of the squeezing parameters $\chi_i$ and the rotation angle $\theta$ for both (i)~one-axis twisting, and (ii) two axis twisting.}
\end{figure}

\begin{figure}[!h]
\centering
    \includegraphics[width=0.6\textwidth]{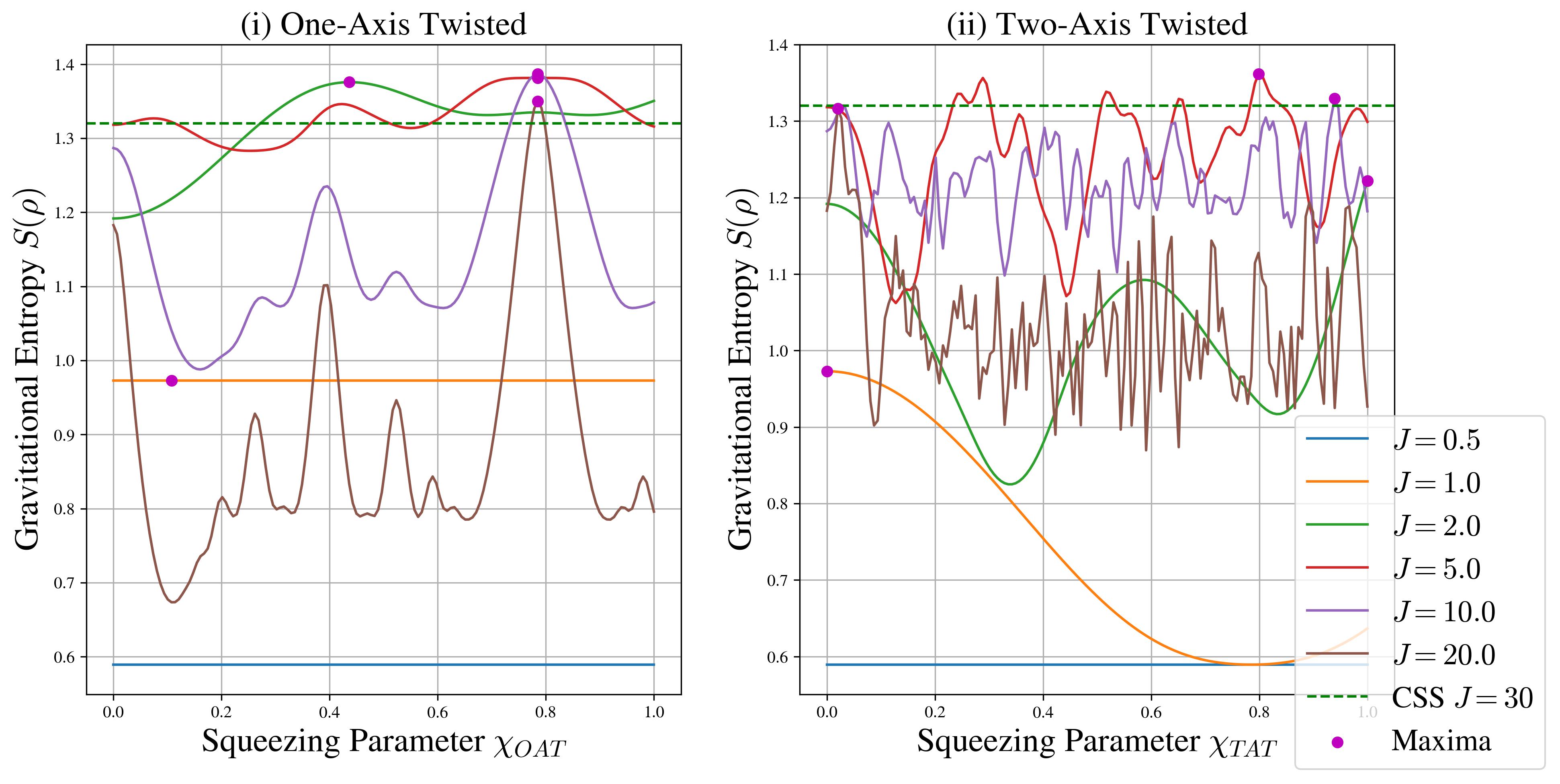}
    \caption{\label{plt:squeezed_1}  Gravitational Entanglement as a function of the squeezing parameter $\chi_i$ of CSS input state with $\theta = \pi/2$, for both (i) one-axis twisting and (ii) two axis twisting.}
\end{figure}
\end{samepage}

\newpage

\section{Negativity Numerical Results}\label{app:negativity}

In this Appendix, the results regarding the negativity generated by the gravitational interaction for different spins lengths and states are reported. For the related discussion, see Sc. \ref{sec:witness}.

\begin{figure}[!h]
  \centering
  \subfloat[J = 5]{\includegraphics[width=0.3\textwidth]{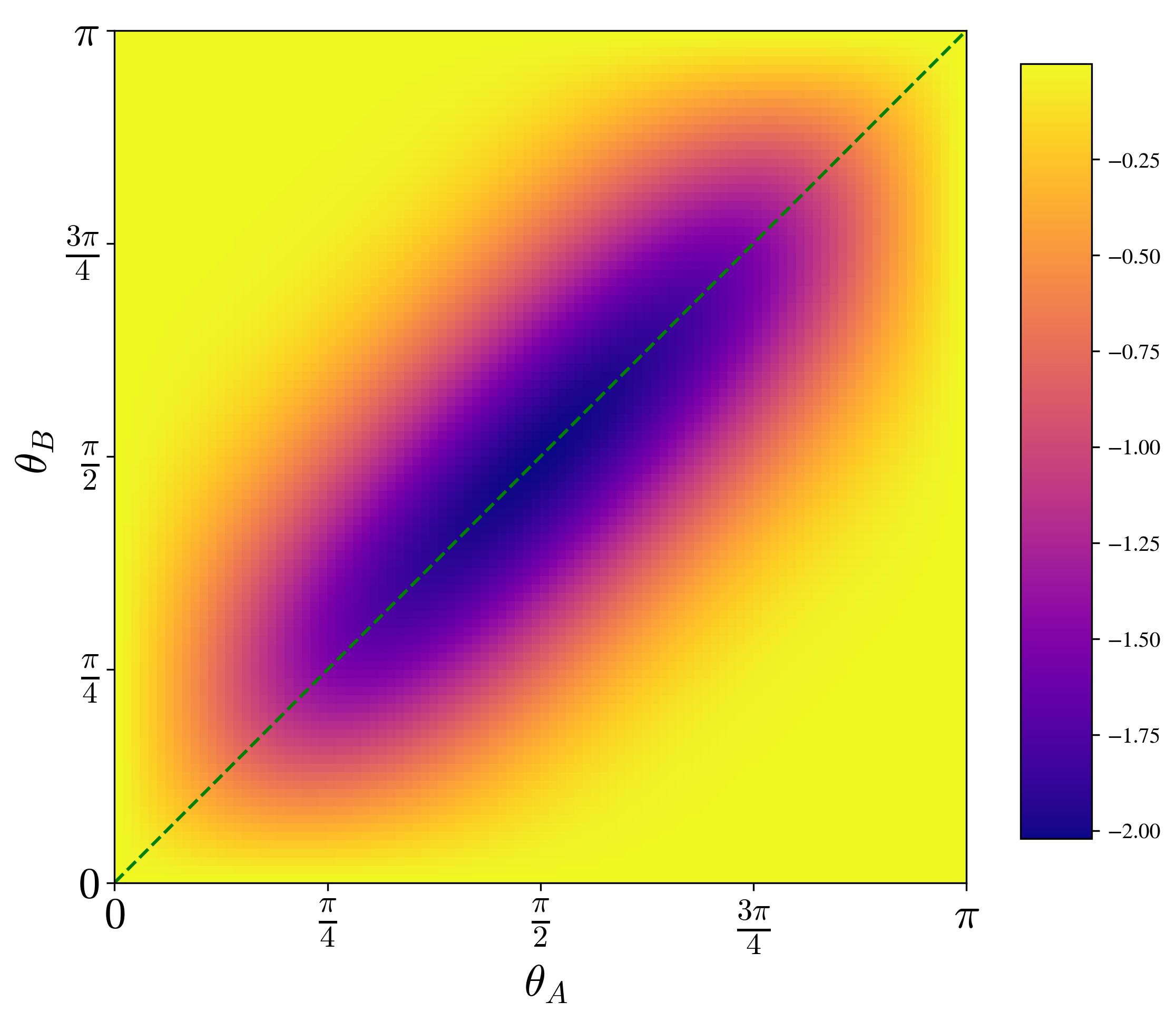}}
  \subfloat[J = 10]{\includegraphics[width=0.3\textwidth]{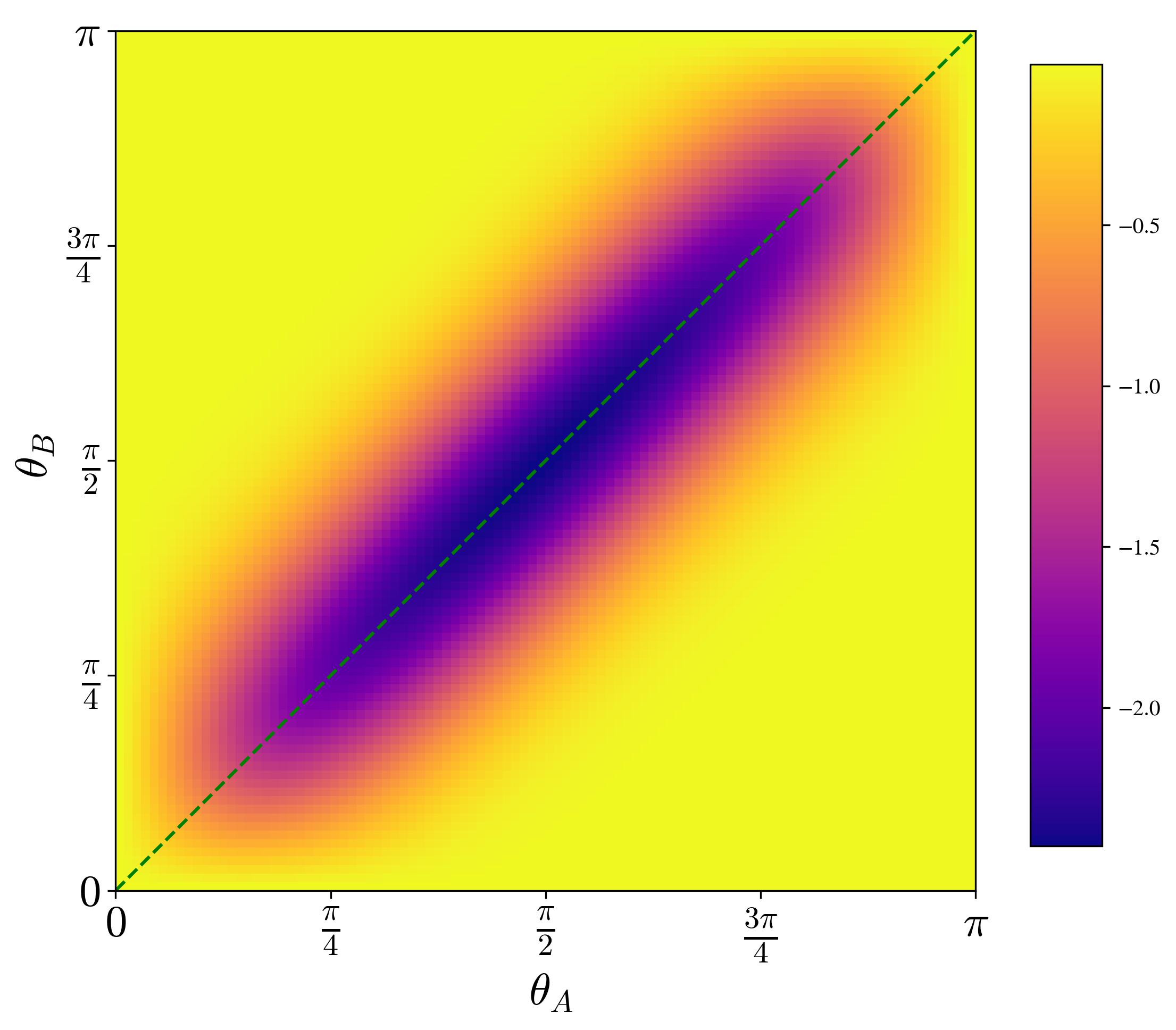}}
  \caption{Negativity for Spin Coherent States with two angles parameters}\label{nega_thetas}
\end{figure}

\begin{figure}[!h]
  \centering
  \subfloat[Negativity as a function of $\theta_A$]{\includegraphics[width=0.31\textwidth]{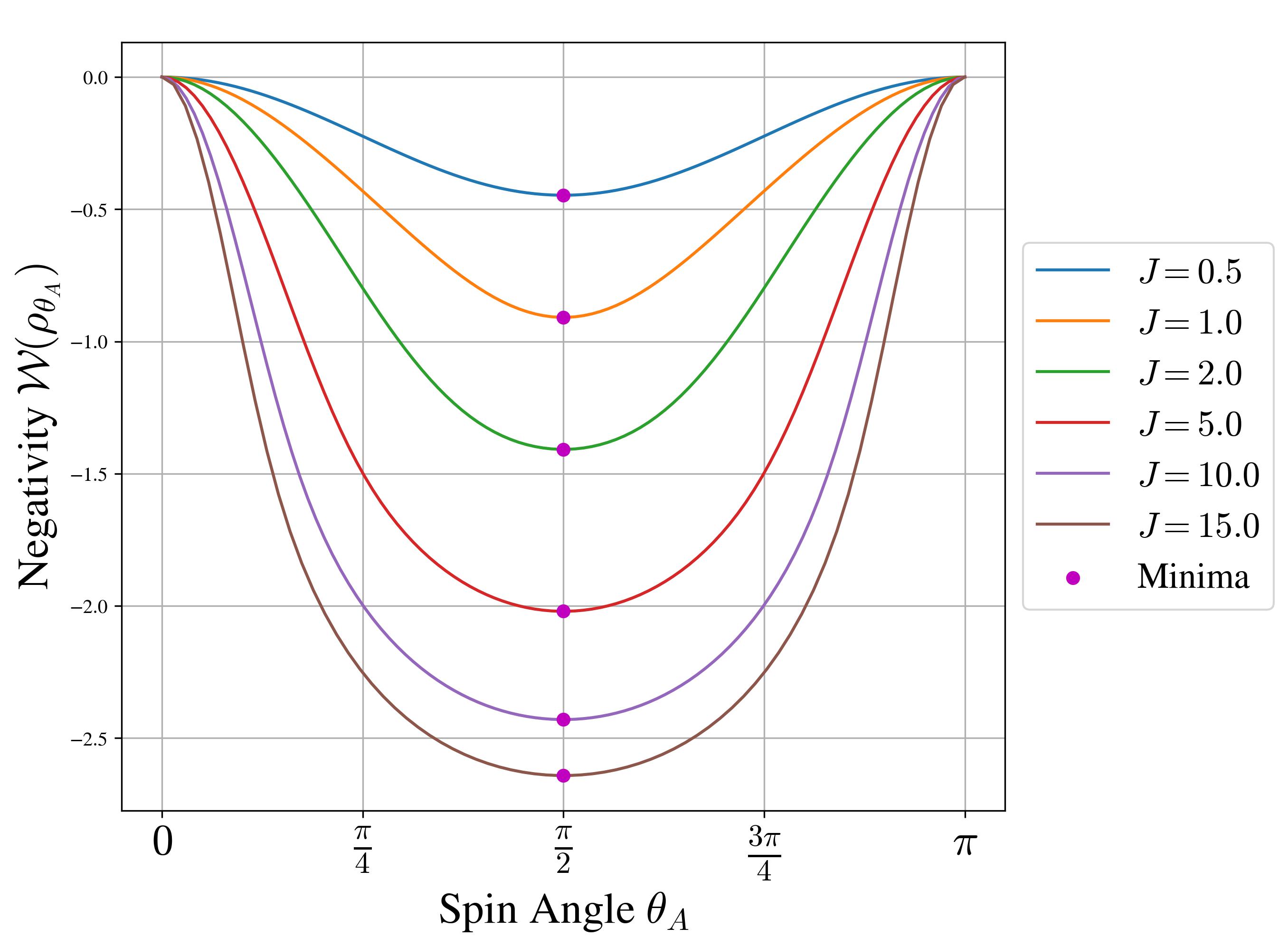}}
  \hfill
  \subfloat[Minimal Negativity as a function of J]{\includegraphics[width=0.33\textwidth]{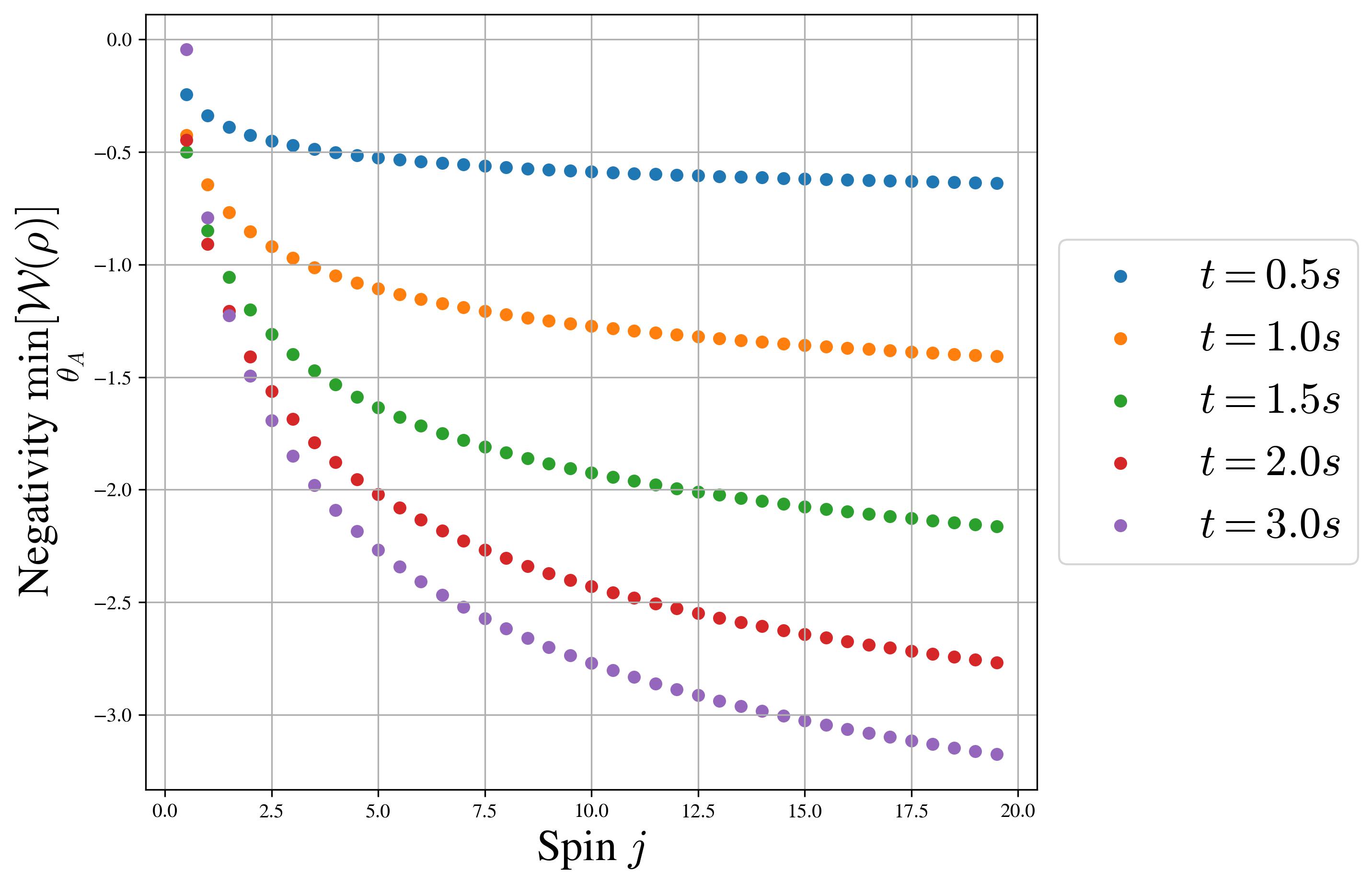}}
  \hfill
  \subfloat[Minimal Negativity as a function of $\tau $]{\includegraphics[width=0.31\textwidth]{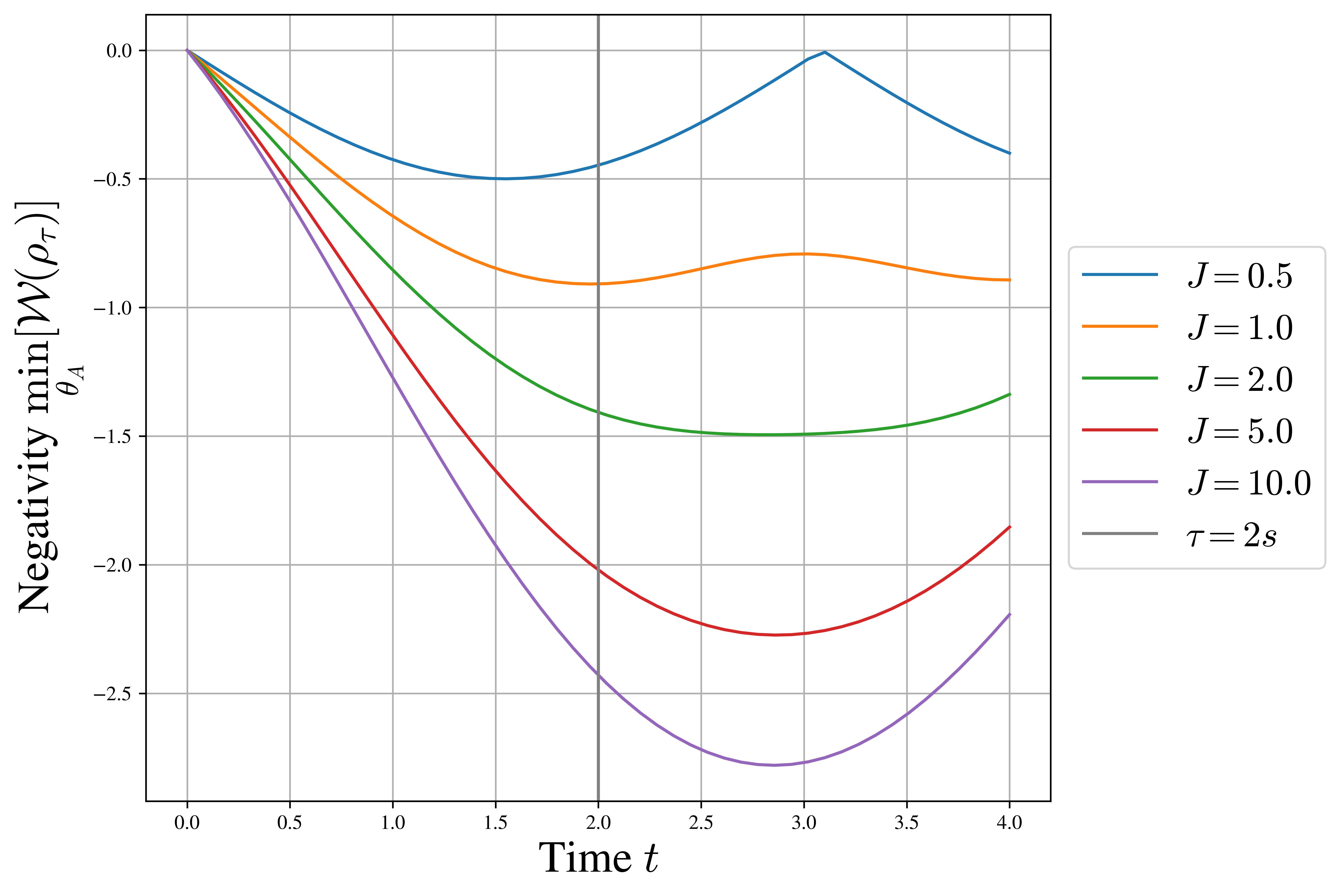}}
  \caption{Negativity for Spin Coherent States with one angles parameters}
\end{figure}

\begin{figure}[!h]
  \centering
  \subfloat[Two-parameters Negativity: J = 5]{\includegraphics[width=0.31\textwidth]{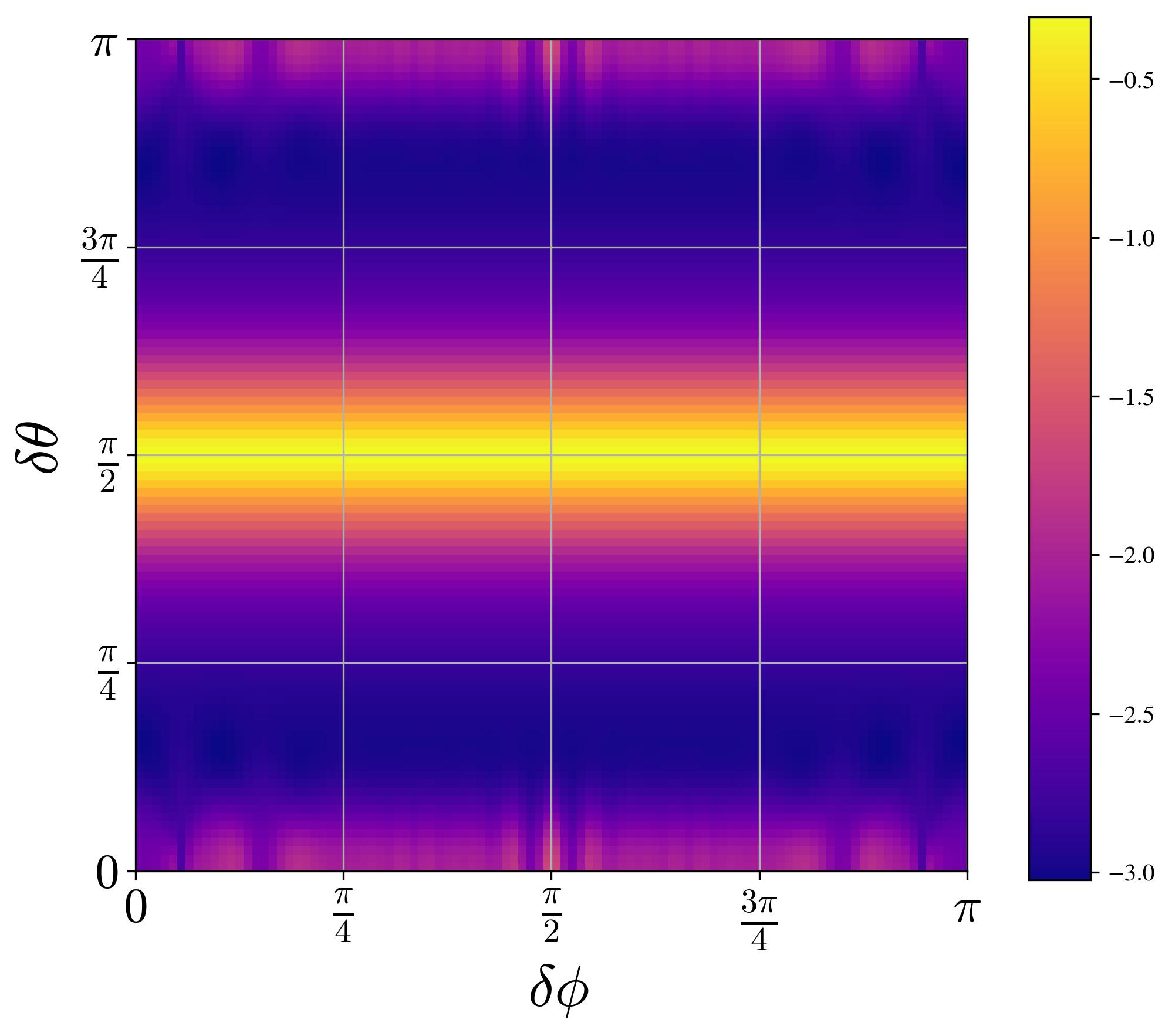}}
  \hfill
  \subfloat[Two-parameters Negativity: J = 10]{\includegraphics[width=0.31\textwidth]{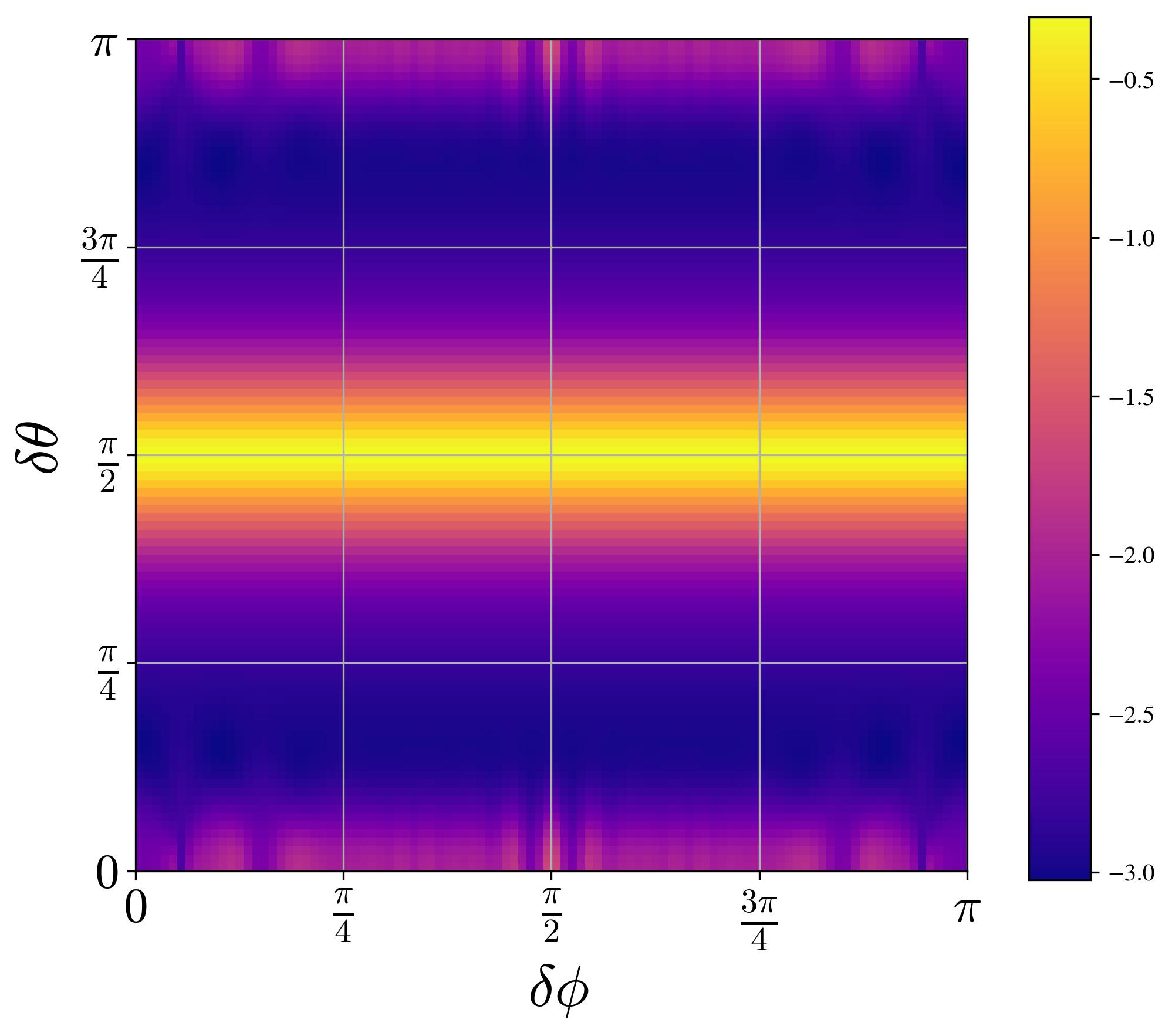}}
  \hfill
  \subfloat[One-parameter Negativity:]{\includegraphics[width=0.31\textwidth]{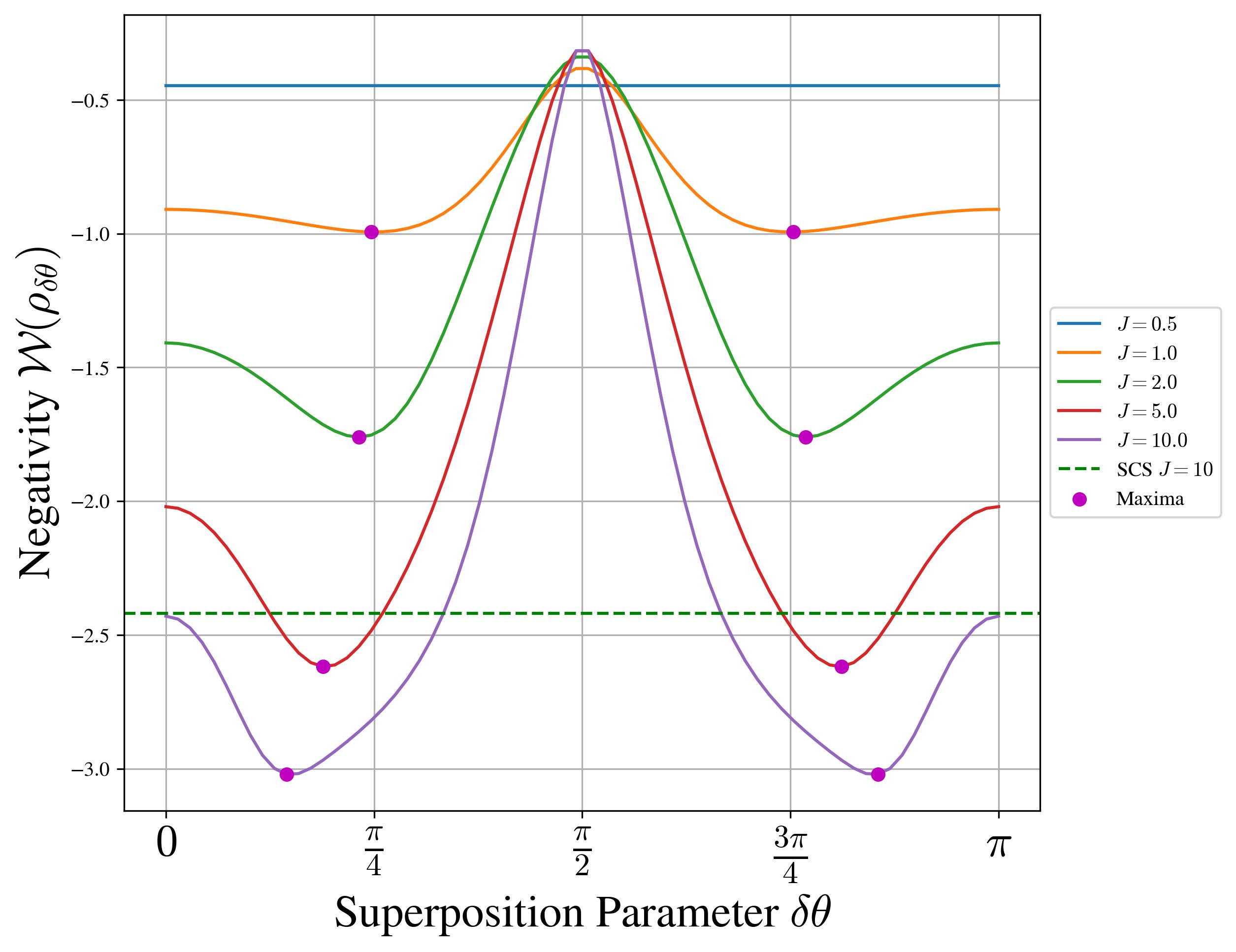}}
  \caption{Negativity for Superposition of CSS, as function of the two parameters $\delta \phi$ and $\delta \theta$ for (a) $J=5$, and (b) $J=10$, (c) and as function of $\delta \theta$ for different spins values.}
\end{figure}

\newpage

\section{\label{app:dechoerecce}Including Decoherence in the Model and Sources of Decoherence}

The general evolution due to decoherence caused by random particle scattering on a system in a superposition is described by the Master equation:~\cite{schlosshauer_quantum_2007,schlosshauer_quantum_2019},
\begin{equation*}
\label{deco_evolution}
    \frac{\partial \rho (\mathbf{x}, \mathbf{x}', t)}{\partial t} = - F(\mathbf{x}-\mathbf{x}') \rho (\mathbf{x}, \mathbf{x}', t)
\end{equation*}
where $F$ plays the role of decoherence factor and, generally, it is given by:~\cite{schlosshauer_quantum_2007,schlosshauer_quantum_2019}
\begin{equation}\label{eq:F-factor}
    F(\mathbf{x}-\mathbf{x}') = \int d q \rho(q) v(q) \int \frac{d \hat{\mathbf{n}}d \hat{\mathbf{n}}'}{4  \pi} \left( 1 - e^{iq (\hat{\mathbf{n}} - \hat{\mathbf{n}}')(\mathbf{x} - \mathbf{x}')/\hbar } \right) \left| f(q \hat{\mathbf{n}}, q \hat{\mathbf{n}}' ) \right|^2
\end{equation}
where $q$ is the normalized momentum of the incoming particle, $\rho(q)$ the momentum-space density, $v(q)$ the speed of particles with momentum $q$, $f(q \hat{\mathbf{n}}, q \hat{\mathbf{n}}' )$ the scattering amplitude\footnote{Defined according to: $\braket{q| \hat{T}| q' } = \frac{i }{2 \pi \hbar q } \delta \left(q- q' \right) f(\mathbf{q} , \mathbf{q}' )$. With $T$ the scattering part of the $S$-matrix.}. 
It is possible to simplify the above equation taking the short-wavelength limit, where the wavelength of the environmental particles $\lambda$ is much smaller than the superposition size, $\lambda \ll \Delta x$ (using the expression of the wavelength in terms of its momentum this is equivalent to $ \frac{q |\mathbf{x} - \mathbf{x}'| }{h} \gg 1$), or the long-wavelength limit, where $\lambda\gg\Delta x$ (or equivalently $ \frac{q |\mathbf{x} - \mathbf{x}'| }{h} \ll 1$).
Within these limits, the integrals over $\mathbf{n}, \mathbf{n}'$ in Eq.~\ref{eq:F-factor} can be simplified and evaluated. The evolution of the off-diagonal terms of the density matrix can be described in the short-and long-wavelength limit, respectively, as:~\cite{schlosshauer_quantum_2007}
\begin{equation*}
    \frac{\partial \rho (\mathbf{x}, \mathbf{x}', t)}{\partial t} = - \gamma_{\text{short}} \rho (\mathbf{x}, \mathbf{x}', t) \, ,\;\;\;\;\;\;\;   \frac{\partial \rho (\mathbf{x}, \mathbf{x}', t)}{\partial t} = - \Gamma_{\text{long}} (\mathbf{x} - \mathbf{x}')^2 \rho (\mathbf{x}, \mathbf{x}', t) \, .
\end{equation*}
where $\gamma_{short} $ and $\Gamma_{long}$ are decoherence rates (the former in units of Hz, while the latter of Hz/m$^2$), and are computed via scattering theory according to the process that they describe. In order to compare how these processes effect the negativity for different spins values, the decoherence rates will be kept as a variable. 

Each subsystem of the bipartite system describing the masses is assumed to evolve under decoherence uncorrelated to the other subsystem, i.e. the scattering particles do not scatter with both masses. Furthermore, it is assumed that the decoherence rate is the same for both subsystems, i.e. both masses are surrounded by the same environment and have the equivalent scattering properties. Under these assumptions and after a time $\tau$, the density matrix of the composite system under decoherence in the short-wavelenght limit:
\begin{equation*}
   \rho_{m, m', n, n'}^{(\text{short})} = \rho_{m, m', n, n'} e^{-(2-\delta_{mm'}-\delta_{nn'})\gamma_\text{short}\tau}.
\end{equation*}
For the short-wavelength, and were $m$ and $m'$ ($n$ and $n'$) are the spin quantum number of subsystem A (B). For the long-wavelength:
\begin{equation*}
    \rho_{m, m', n, n'}^{(\text{long})} =  \rho_{m, m', n, n'}e^{- \Gamma_{\text{long}} (\Delta x)^2 \left[ \left(m - m'\right)^2 + \left(n - n'\right)^2  \right] \tau}
\end{equation*}
The decoherence terms cause an exponential decay of the off-diagonal terms of the density matrix. After a certain time all off-diagonal terms will go to zero, and the density matrix becomes a classical mixed state.

It is possible to note that in the spin $1/2$ case, the two processes are described by the density matrix:
\begin{equation*}
    \rho_{1/2}^{(\text{short/long})} = 
    \begin{pmatrix}
    \rho_{00,00}       & \rho_{00,10} e^{- \gamma \tau} & \rho_{10,00} e^{- \gamma \tau}  &  \rho_{10,10} e^{- 2 \gamma \tau} \\
    \rho_{00,01}      e^{- \gamma \tau}   & \rho_{00,11} & \rho_{10,01}  e^{- 2 \gamma t}  &  \rho_{10,11} e^{- \gamma \tau}  \\
    \rho_{01,00}       e^{- \gamma \tau}  & \rho_{01,10} e^{- 2 \gamma \tau}  & \rho_{11,00} &  \rho_{11,10} e^{- \gamma t}  \\
    \rho_{01,01}   e^{- 2 \gamma \tau}    & \rho_{01,11}  e^{- \gamma \tau}  & \rho_{11,01} e^{- \gamma \tau}  &  \rho_{11,11} 
\end{pmatrix}
\end{equation*}
where it was possible to define a total decoherence rate $\gamma_{\text{tot}} = \gamma_{ \text{short}} + \Gamma_{ \text{long}} \Delta x^2$. This will not be possible in the case of spins with $J\geq 1/2$.

\end{document}